\documentclass[twocolumn,showpacs,preprintnumbers,amsmath,amssymb]{revtex4-1}

\usepackage{graphicx}
\usepackage{dcolumn}
\usepackage{amsmath,mathrsfs,amssymb,amsthm}
\usepackage{hhline}
\usepackage{wasysym,enumerate,color}
\usepackage{epstopdf}
\usepackage{verbatim}
\usepackage{hyperref}
\usepackage{color}
\usepackage{ulem} 

\renewcommand{\emph}[1]{\textit{#1}} 

\definecolor{darkgreen}{rgb}{0,0.5,0}
\definecolor{purple}{rgb}{0.35,0,0.35}
\definecolor{orange}{rgb}{1,0.5,0}
\definecolor{darkred}{rgb}{.7,0,0}
\definecolor{darkblue}{rgb}{0,0,.3}
\definecolor{grey}{rgb}{.6,.6,.6}
\definecolor{dimgreen}{rgb}{0.2,0.6,0.1}
\hypersetup{colorlinks,linkcolor=blue,urlcolor=blue,citecolor=blue}

\newcommand{\be}{\begin{equation}}
\newcommand{\ee}{\end{equation}}
\newcommand{\bea}{\begin{eqnarray}}
\newcommand{\eea}{\end{eqnarray}}
\newcommand{\rmd}{{\rm d}}

\newcommand{\bS}{{\bf S}}

\newcommand{\G}{{\Gamma}}
\newcommand{\g}{{\gamma}}

\newcommand{\cG}{{\cal G}}

\newcommand{\e}{\varepsilon}
\newcommand{\sgn}{{\rm sgn}}
\newcommand{\w}{\omega}
\newcommand{\s}{\sigma}

\newcommand{\down}{\downarrow}

\begin{document}

\title{Transport in a hybrid normal-topological superconductor Kondo model}

\author{Razvan Chirla}
\affiliation{Department of Physics, University of Oradea, 410087, Oradea, Romania}

\author{I. V. Dinu}
\affiliation{National Institute of Materials Physics, POB MG-7, 77125 Bucharest-Magurele, Romania}

\author{V. Moldoveanu}
\affiliation{National Institute of Materials Physics, POB MG-7, 77125 Bucharest-Magurele, Romania}

\author{C\u at\u alin  Pa\c scu Moca}

\affiliation{BME-MTA Exotic Quantum Phase Group, Institute of Physics, Budapest University of Technology and Economics, H-1521 Budapest, Hungary}

\affiliation{Department of Physics, University of Oradea, 410087, Oradea, Romania}

\date{\today}

\begin{abstract}

We investigate the equilibrium and non-equilibrium transport through a quantum dot in the Kondo regime, embedded 
between a normal metal and a topological superconductor supporting Majorana bound states at its end points. 
We find that the Kondo physics is significantly modified by the presence of the Majorana modes. 
When the Majorana modes are coupled, aside from the Kondo scale $T_K$,  a new energy scale $T^*\ll T_K$ emerges, 
that controls the low energy physics of the system. 
At low temperatures, the ac-conductance is suppressed for frequencies below $T^*$, while the noise spectrum acquires a $\sim \omega^3$ dependence.
At high temperatures, $T \gg T_K$, the regular logarithmic dependence in the
differential conductance is also affected. Under non-equilibrium conditions, and  
in particular in the $\{T, B\}\to 0$
limit, the differential conductance becomes negative. 
These findings indicate that 
the changes in transport may serve as clues for detecting the Majorana bound states in such systems. 
In terms of methods used, we characterize the transport by using a combination of perturbative and renormalization group approaches.

\end{abstract}

\pacs{73.63.Kv, 72.10.Ad, 73.23.Hk, 85.35.Be, 85.75.Ad}

\maketitle

\section{Introduction} \label{sec:Introduction}
Topological phases supporting Majorana fermions constitute an active topic
of current research in the condensed matter community, motivated by the pursuit of
fault-tolerant topological quantum computation~\cite{Alicea.12}.
In general, Majorana bound states
(MBS) appear as zero energy electron-hole states of a superconductor
with broken spin degeneracy. At present, their existence has been predicted in a variety 
of systems, such as $p$-wave spinless superconductors~\cite{Kitaev.01},
three and two-dimensional topological insulators in proximity to
a superconductor~\cite{Fu.08,Fu.09}, the
$\nu=5/2$ fractional quantum Hall state~\cite{Moore.91, DasSarma.05}, 
cold atom wires~\cite{Jiang.11}, optical lattices~\cite{Sato.09}, quantum dot chains~\cite{Fulga.13},
superconductor - double quantum dot systems~\cite{Wright.13}, and 
 helical Shiba chains~\cite{Nadj-Perge.13, Yazdani.97,Poyhonen.14}. It was recently proposed that 
MBS can be engineered by inducing superconductivity in a semiconducting
nanowire with strong spin-orbit coupling~\cite{Lutchyn.10, Oreg.10, Brouwer.12}.
When such nanowires
are exposed to a magnetic field, 
they are driven into the topological phase supporting MBS at their end points.

As the MBS modify the transport properties of the nanowire device, 
several experimental detection schemes were proposed. Tunneling spectroscopical measurements of the zero bias anomaly could 
support the
hypothesis of  MBS~\cite{Mourik.12, Das.12, Deng.12}, 
while other do not rule out alternative explanations, such as the Kondo effect~\cite{Churchill.13}, 
and emerging subgap states~\cite{Lee.12b}.
Alternatively, at equilibrium, in a Josephson junction, the presence of the MBS enables not just the tunneling of Cooper 
pairs but also of single electrons, amounting to a doubling of the periodicity of 
the Josephson energy~\cite{Beenakker.13, Law.11, Ioselevich.11,Rokhinson.12}. Measurement of nonlocal
noise cross correlations of two well-separated dots mediated by MBS was also proposed as a test~\cite{Lu.12,Zocher.13,Li.14}. 
 In other approaches, a quantum dot coupled to two normal leads and to 
a topological superconductor supporting MBS would show dynamical electrical features that could confirm the presence of 
MBS~\cite{Liu.11,Zitko.11,Lee.13}. Finally, other proposals include the measurement of the anisotropic 
susceptibility of a probe impurity~\cite{Shindou.10,Zitko.11b},
interferometric
schemes~\cite{Grosfeld.11,DasSarma.06,Akhmerov.09}, and spin-polarized scanning tunneling microscopy~\cite{Sticlet.12}.
Although these investigations have
made important steps toward the observation of MBS, a clear fingerprint that 
can provide conclusive evidence for the existence of MBS is still missing. 

\begin{figure}[!tbhp]
\begin{center}
\includegraphics[width= 1.0\columnwidth]{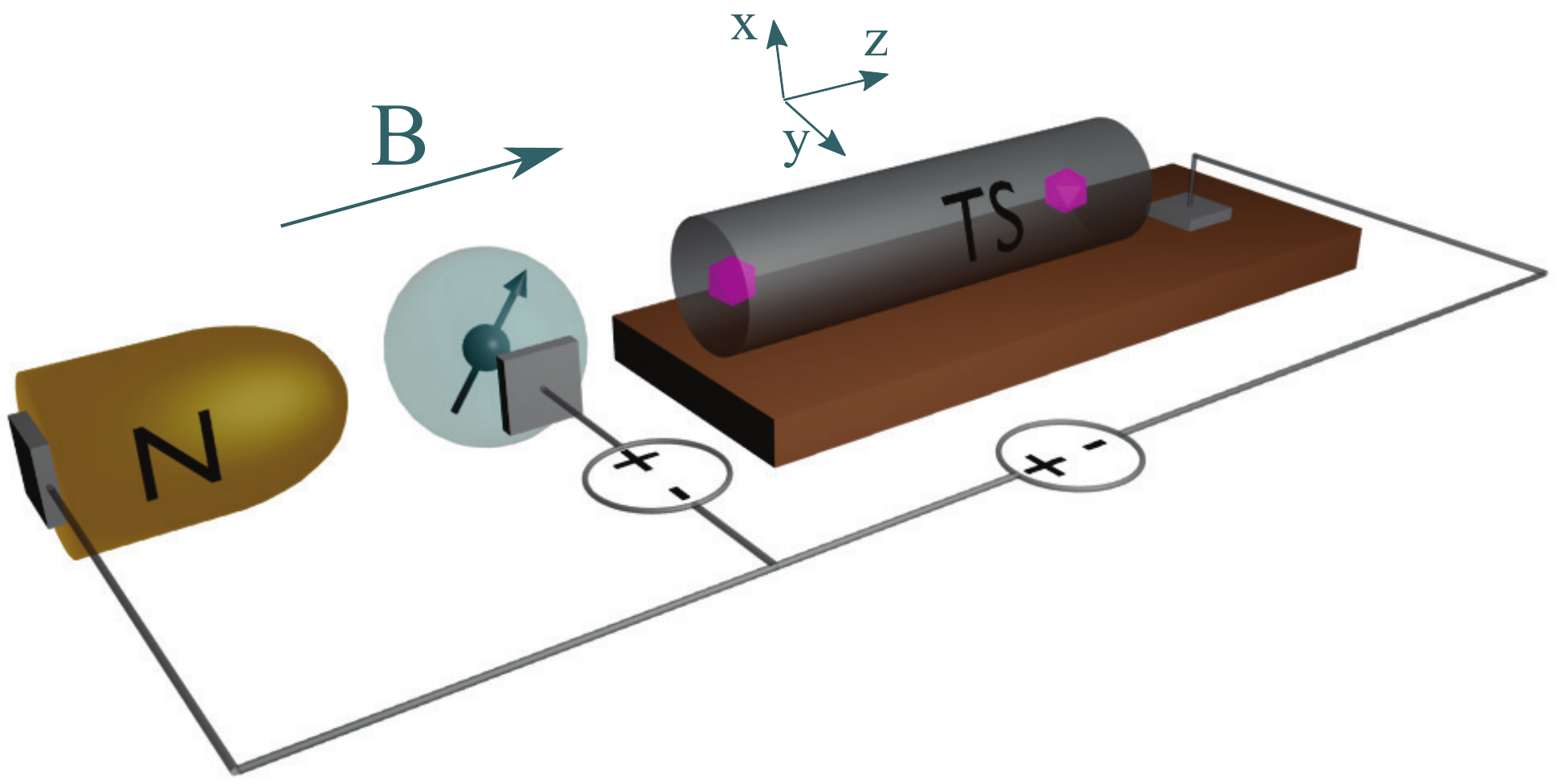}
\end{center}
\caption{	
(Color online) Sketch of the setup to measure the conductance and noise. 
The dot is in the Kondo regime and is occupied by a single spin $S=1/2$.  
The left lead (N) is a normal metal, while the right one (TS) 
is a topological superconductor. 
}\label{fig:sketch}
\end{figure}

The purpose of this work is to provide a general analysis
of the transport across a quantum dot (QD) embedded  
between a normal metal and a nanowire-type topological 
superconductor~\cite{Golub.11,Leijnse.11,Cheng.13,Leijnse.14}, as presented in ~Fig.~\ref{fig:sketch}. 
As the transport and spectroscopic measurements are the most accessible ones, we shall try to 
identify possible changes in transport coming from the interplay with the Majorana modes.
Throughout this work we shall focus on the situation when the dot is in the local moment regime, 
described in terms of the local spin $\bS$. When the charge fluctuations are neglected, 
the exchange coupling between the dot and the normal metal (NM) lead is given by the Kondo interaction
\begin{equation}\label{eq:H_Kondo}
H_{\rm int}^{NM} = {J \over 2}\sum_{\s\s'}  \, 
\psi_{\s}^{\dagger}\,  \boldsymbol{\s}_{\s\s'} \psi_{\s'}\, \mathbf{S}\, ,  
\end{equation}
with $J$ the Kondo coupling, which in general controls the Kondo temperature 
\begin{equation}
T_K\approx D\, e^{-1/(\varrho_0 J)}.
\label{eq:T_K}
\end{equation}
Here, $\varrho_0$ is the density of states for the electrons in the normal electrode, and $2D$ is the bandwidth
of the conduction electrons. 
Although Eq.~\eqref{eq:T_K} defines the Kondo scale in the absence of the MBS, we shall prove in 
Sec.~\ref{sec:RG} that this definition indeed extends to the case when the TSC is coupled to the dot, 
 and that $T_K$  can be defined only in terms of the dimensionless coupling $j = \varrho_0\, J$. 
In Eq.~\eqref{eq:H_Kondo}, the field operator
\begin{equation}
\psi^{\dagger}_{\s} = \sqrt{\varrho_0}\int_{-D}^{D} c_{\s}^{\dagger}(\e) d\e\, ,  
\end{equation}
creates an electron with spin $\s$ in the normal lead, and $\boldsymbol{\s}$ are the Pauli matrices. 
While the Kondo screening of the dot spin by the conduction electrons of two normal contacts is 
well understood by now~\cite{Pustilnik.04}, it is interesting to understand how the Kondo  correlations develop when a 
normal metal lead is replaced by a topological superconductor. It is known that
in a NM-QD-Superconductor setup~\cite{Oguri.04, DeFranceschi.10}, the Kondo correlations 
and the superconducting correlations controlled by $\Delta$, the superconducting gap, compete and influence 
the ground state of the magnetic moment. As a consequence, 
 a quantum phase transition emerges when $T_K\approx \Delta$. On one side of the transition point, when $T_K\gg \Delta$, a
 singlet ground state is formed, as the local spin becomes screened, while in the opposite limit, $\Delta\gg T_K$, the ground state becomes a doublet,
 as the spin does not couple to the environment.  
 
 Here we are interested in what happens to the Kondo correlations and to the transport across the dot 
 in a slightly different configuration, 
 when  a normal lead is replaced 
 with a nanowire in the topological superconducting (TS) phase. 
For a non-interacting wire, the TS phase emerges when  $h>\sqrt{\Delta^2+\mu^2}$~\cite{Sato.09}, where $\Delta$
is the pairing amplitude inherited from the nearby superconductor, $\mu$ is the 
chemical potential and $h$ is the Zeeman energy of the external magnetic field. 
In the presence of strong interactions, it was recently 
suggested \cite{Stoudenmire.11} that such 
systems support Majorana fermions even in the absence of a magnetic field.
When the gap is large, as we are interested in the low energy physics, $E\ll \Delta$, 
it is a reasonable approximation to model the TS phase simply by a pair of
MBS formed at the ends of the wire~\cite{Golub.11}. In this case, 
the length of the nanowire influences the hybridization $\xi$ of the two Majorana modes. 
 It is exponentially small 
and depends on the length $L$ of the superconductor as $\xi\propto \exp(-L/\zeta)$, 
with $\zeta$ the coherence length of the TSC. 
When $L\gg \zeta$, the MBS states become decoupled, since  $\xi=0$. 

The interplay between the Kondo and Majorana physics is analyzed 
by constructing an effective Kondo Hamiltonian, in the spirit 
of the Schrieffer-Wolff transformations~\cite{Schrieffer.66},  from the single impurity Anderson model 
described by the Hamiltonian ~\eqref{eq:H_Anderson} in Appendix \ref{app:SW}.
By projecting out the empty and doubly occupied states of the dot, an effective 
Hamiltonian of the form \eqref{eq:H_int} is obtained.
We consider the configuration in which the
normal lead is much more strongly coupled to the dot than to the MBS~\cite{Golub.11}, 
which translates in terms of the Kondo couplings
to $J\gg J_M$ in Eq.~\eqref{eq:H_int}.  

In the present work we shall characterize the equilibrium (i.e., zero voltage) and non-equilibrium  
conductance, investigate the correlations of the current $I(t)$ across the dot in time
\begin{equation}
S(t-t')=\frac{1}{2}\langle \{I(t), I(t')\}\, \rangle, 
\label{eq:noise}
\end{equation} 
and determine the noise spectrum. In terms of the methods used, 
the analysis of the equilibrium ac-transport is carried out by
using the numerical renormalization group (NRG) approach~\cite{Wilson.75, Krishnamurthy.80, Bulla.2008}.
To obtain a clear picture, in the weak coupling regime, $\{\omega, T\}\gg T_K$, 
we supplement our numerical results 
with analytical perturbative and renormalization group calculations.
To  address the out-of-equilibrium situation, the 
current-voltage characteristics are obtained 
by using a modified version of the real time functional renormalization group (fRG) approach on the 
Keldysh contour, developed in Ref.~\cite{Rosch.05, Moca.14}. 
\begin{figure}[!tbhp]
\begin{center}
\includegraphics[width= 1\columnwidth, clip]{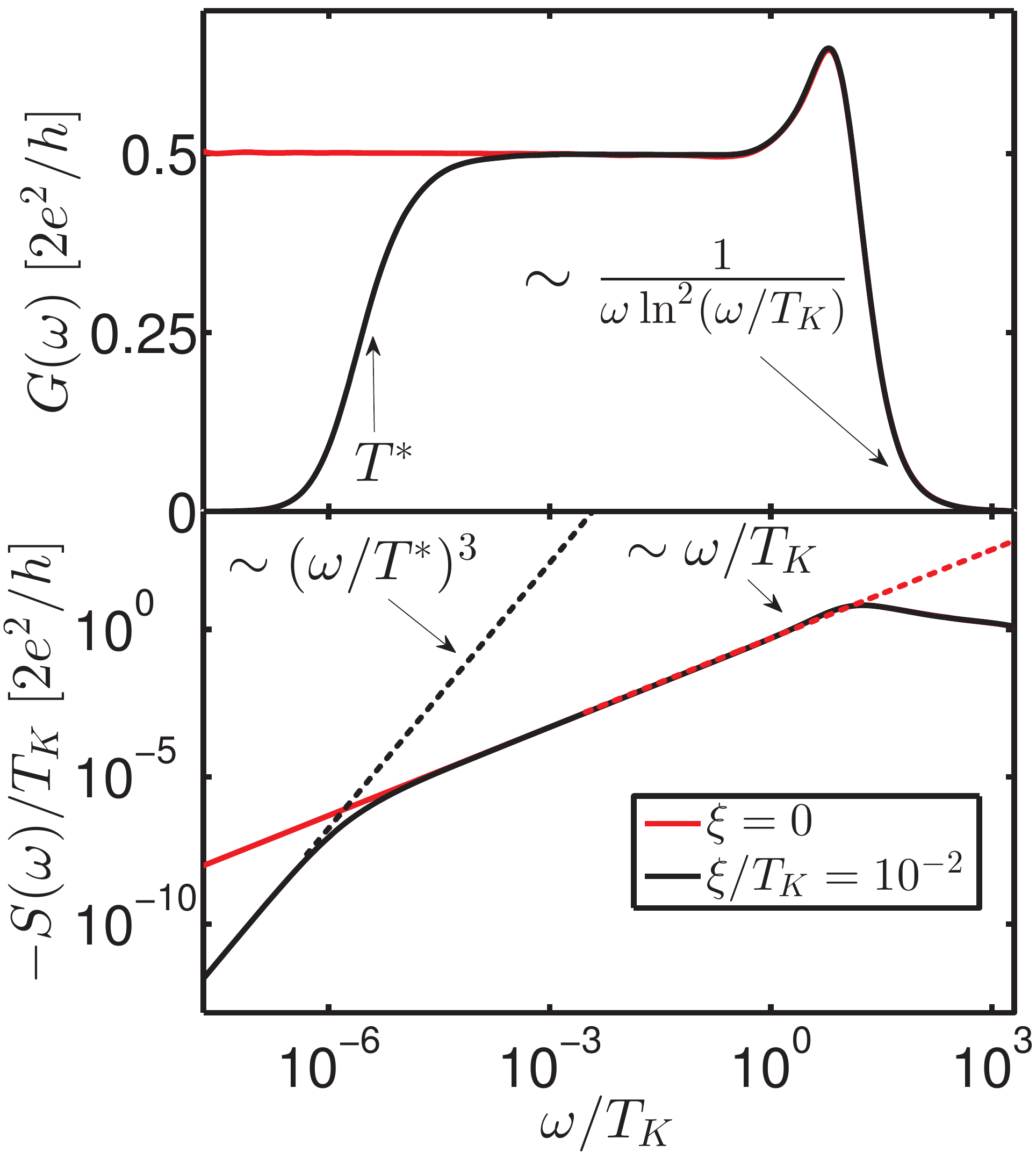}
\end{center}
\caption{\label{fig:G_omega}	
(Color online) Frequency dependent conductance $G(\omega)$ and current noise $S(\omega)$ at $T=0$ 
for $\xi=0$ and $\xi\ne 0$. The characteristic scale $T^*$ 
is a manifestation of the parity conservation and 
can be associated with the energy below which the cotunneling events become 
suppressed.}
\end{figure}
As we shall see, the ac-conductance and noise spectra display a very rich structure. When 
$\xi \ne 0$, as shown in Fig.~\ref{fig:G_omega}, below the Kondo temperature $T_K$, a new energy scale $T^*$ emerges,
below which the conductance vanishes~\cite{Leijnse.11}.
As we shall explain below,  
the inelastic cotunneling processes become suppressed below $T^*$ when the even-odd 
degeneracy of the Hamiltonian sectors is broken by $\xi\ne 0$. Furthermore, in this regime,  the noise spectrum scales as
\begin{equation}\label{eq:noise_scaling}
\frac{S(\omega)}{T^*} \approx  -\alpha\, \Big (\frac{e^2}{h}\Big )\,   \Big (\frac{|\omega|}{T^*}\Big )^3+\dots, \;\;\;  \omega \ll T^*\, , 
\end{equation}
with $\alpha$ a constant of the order $\sim 1$.  
This behavior has a simple physical explanation in terms of parity conservation:
In the absence of a parity relaxation mechanism, an electron injected from the normal lead
can give rise with a certain probability to an outgoing electron in the TSC lead. Before such a process can occur,
another electron from the normal lead must be injected, 
which restores the parity of the initial state, and therefore two electrons must be transferred across the dot.   
Thus, at equilibrium, no transport is possible and $G(\omega\to 0)=0$. In the small frequency limit, 
$G(\omega)\sim \omega ^2$, and by the fluctuation-dissipation theorem this implies a $|\omega|^3$ dependence in noise spectrum
in Eq.~\eqref{eq:noise_scaling}. 
 At energies of the order $\omega \sim T^*$, the inelastic cotunneling processes are restored,
and a half unitary conductance emerges.  As a result, the noise acquires a linear dispersion, as given in Eq.
\eqref{eq:noise_linear}. The $S(T^*\ll \omega\ll T_K)\approx \alpha\,\omega +\dots $ dependence is a consequence of the 
Fermi liquid nature of the Kondo state~\cite{Sindel.05}.

The Majorana modes change also the high energy behavior. In the perturbative limit, 
the "regular" Kondo transport is characterized by a logarithmic dependence on energy, as the differential 
conductance is well described by the expression~\cite{Moca.14}
\begin{equation}\label{eq:G_Kondo}
G(\{eV, T, B\}\gg T_K) \propto \left(\frac{e^2}{h} \right) \frac{1}{\ln^2(\{|eV|, T, B \}/T_K)}\, , 
\end{equation}
with $V$ the bias across the dot and $B$ the external magnetic field. This behavior is a hallmark of the Kondo effect,
and the divergence in Eq.~\eqref{eq:G_Kondo} when $T\to T_K$, indicates the breakdown of the perturbative approach. 

A perturbative treatment, backed up by the NRG analysis, indicates that for the hybrid system that we consider, the standard 
Kondo behavior is  
modified, and that the differential conductance shows a much stronger decrease as a function of energy. 
For example, the temperature dependence in $G(T)$ is captured by the expression 
\be
G(T\gg T_K) \approx \left (\frac{e^2}{h} \right) \frac{3\pi^2}{8}\, \frac{1}{\varrho_0\, T} \frac{p^2}{\ln^2(T/T_K)}\, , 
\label{eq:G_pert_T}
\ee 
 with $p$ a simple number giving the asymmetry of the dot, as  defined in Sec.~\ref{sec:Hamiltonian}. Our perturbative result
is in agreement with the one obtained in Ref.~\cite{Golub.11}, where a similar temperature dependence was obtained. 
When compared to the regular Kondo behavior,~\cite{Sindel.05},  
the equilibrium ac-conductance $G(\omega)$, as prescribed by Eq.~\eqref{eq:G_pert},
is also modified. 
(see also Fig.~\ref{fig:G_omega} for the NRG results). As a result, the leading logarithmic corrections for the noise is found to be
\be
S(\omega\gg T_K) \approx \left (\frac{e^2}{h} \right) \frac{3\pi^2}{2\varrho_0}\, \frac{p^2}{\ln^2(|\omega |/T_K)} .
\label{eq:S_pert}
\ee
It indicates 
that the noise decreases with the frequency, which is   strikingly different from the 
the almost linear dependent spectrum, $S(\omega\gg T_K)\sim \omega/\ln^2(|\omega|/T_K)$, that develops in the regular
Kondo model~\cite{Moca.11}.
\begin{figure}[!ptbh]
\begin{center}
\includegraphics[width= 0.9\columnwidth]{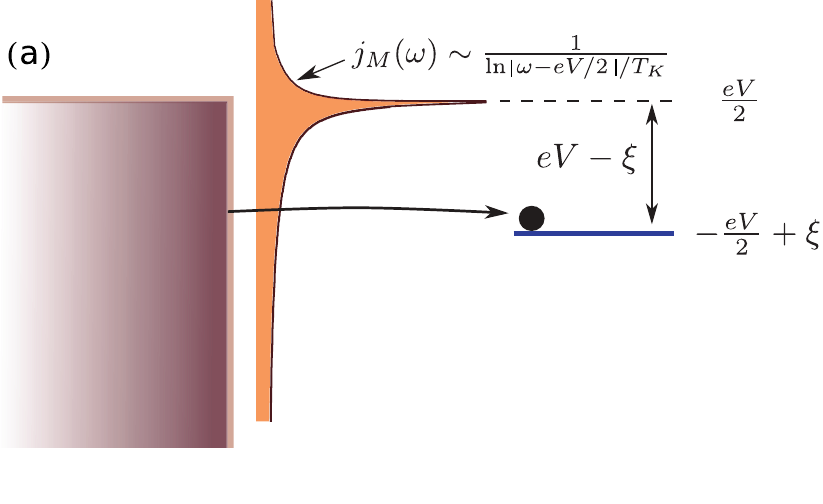}
\includegraphics[width= 0.9\columnwidth]{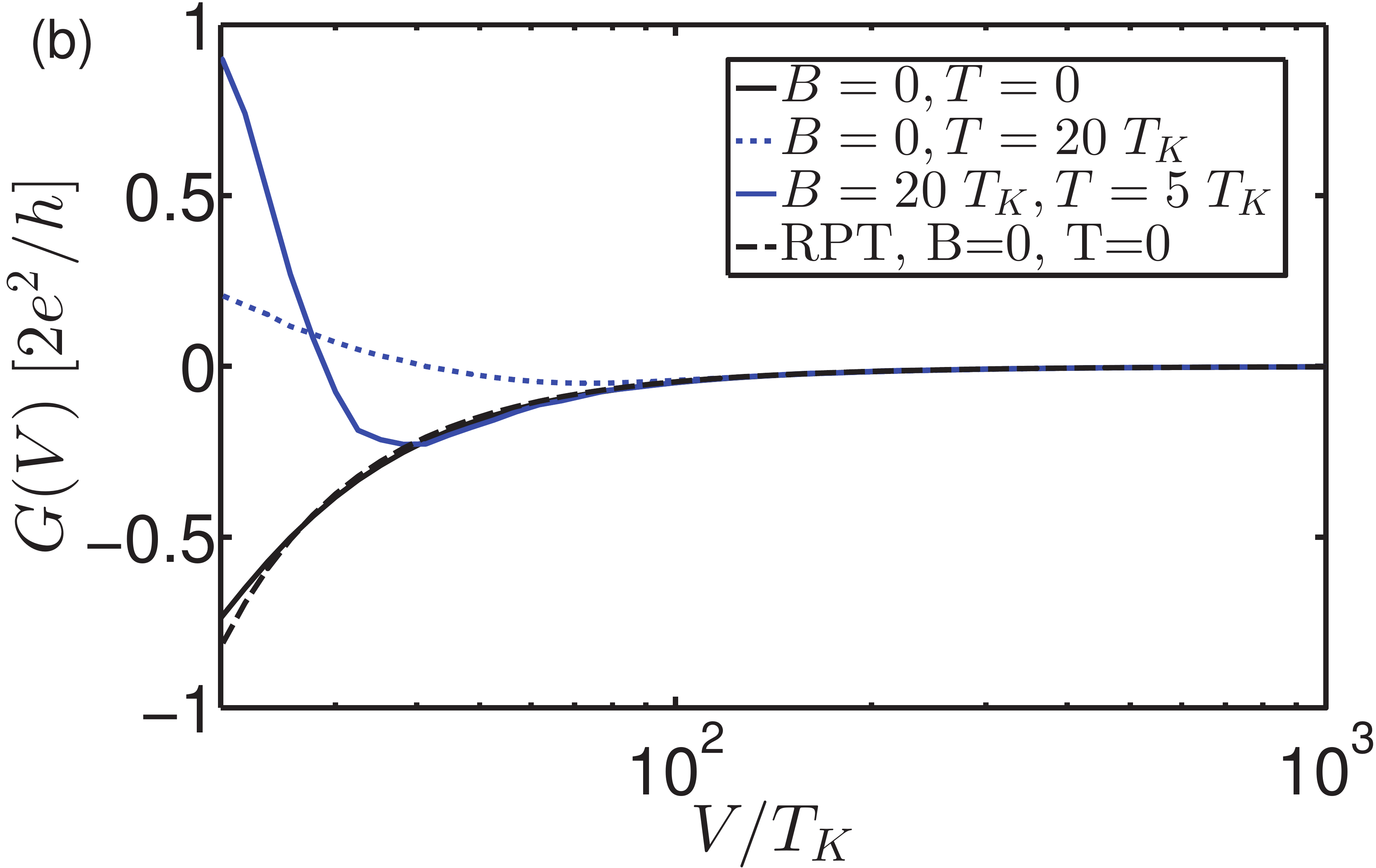}
\end{center}
\caption{(Color online)\label{fig:diff_cond} (a)
 Voltage dependence of  $j_M(eV)$, showing the logarithmic divergence $\sim 1/\ln(|\omega-eV/2|\,T_K)$ pinned at $\sim eV/2$. The arrow
pictures an electronic process that contributes to the current, in which one electron is transferred resonantly from the normal lead to the Majorana bound state. 
(b) The non-equilibrium differential conductance $G(V)$ for different temperatures and magnetic fields. The dashed black 
line represents the renormalized perturbative result. We used the parameters $\xi=0$, $j=0.125$ and $j_M/j=p=0.1$.
}
\end{figure}

So far we have restricted our discussion to the equilibrium situation. The description of the 
non-equilibrium problem is in general difficult since the perturbation approach fails, 
as a resummation to infinite order of the most diverging diagram is in general needed. In that regard,
in Ref.~\cite{Moca.14} we have developed a real time function renormalization group  (fRG) formalism that 
provides an elegant solution for investigating the non-equlibrium transport.  
The method is related in spirit with the other fRG approaches developed in 
Refs.~\cite{Pletyukhov.12,Metzner.12}.
In the present work we extend the formalism to the setup configuration in Fig.~\ref{fig:sketch}.  
Like all the other RG approaches, the fRG is suitable only in the weak coupling regime, where $\max\{eV, T, B\}\gg T_K$. 
At present,  we are not aware of any trustable method that can access the strongly coupled limit at non-equilibrium, 
and this is the reason why the non-equilibrium configuration is investigated only in the perturbative limit. In the fRG
approach, the effective interactions become local in time, and they satisfy an integro-differential set of equations, as
shall be discussed in more detail in Sec.~\ref{sec:RG}. Here, we just want to point out that the dimensionless renormalized 
exchange interaction, 
$j_M(\omega)=\rho_0J_M$, that couples the dot to the MBS, and defined in Eq.~\eqref{eq:couplings_def}, 
displays a singularity that is in resonance with the 
 Fermi energy of the normal lead,  without any other particular features, 
 $j_M(\omega) \sim 1/\ln(|\omega-eV/2|/T_K)$. The upper panel 
 of Fig.~\ref{fig:diff_cond} presents a sketch with the frequency dependence of $j_M(\omega)$, while the exact result
 is presented in Figs.~\ref{fig:sc_eq_B_0} and ~\ref{fig:sc_eq_B}.
 However, the most astonishing result is presented in Fig.~\ref{fig:diff_cond} (lower panel): 
 In the $\{B, T\}\to 0$  limit, the \emph{differential conductance becomes negative}. This behavior can be understood in terms 
 of the resonant transmission between the normal lead and the MBS.  
It is shown in  Eq.~\eqref{eq:noneq_curr}, assuming $\xi=0$, that up to a prefactor, the average current $\langle I\rangle$
is simply given by $\langle I\rangle \propto |j_M(-eV/2)|^2 \tanh(eV/2T)$. At $T=0$, 
the leading logarithmic correction (the first term in \eqref{eq:GV}) vanishes, and 
only the sub-logarithmic correction (second term in \eqref{eq:GV}) survives. 
Then, the  differential conductance becomes negative.  At intermediate voltages and when $T>0$,  
the leading contribution 
becomes dominant and restores the differential conductance to a positive value. 
On the other hand, at finite temperature, the $\tanh$ function has a finite derivative, which can reverse the slope of the current.

The emergence of the new energy scale $T^*$, together with the results for the conductance and noise presented in
Eqs.~\eqref{eq:noise_scaling},~\eqref{eq:G_pert_T} and \eqref{eq:S_pert}, and  
 the prediction of a negative differential conductance under 
non-equilibrium conditions,  are the main results of this paper.  They  pinpoint significant changes 
in the transport across a dot in the Kondo regime as a consequence of the presence of the Majorana modes.  

The rest of  the paper is organized as follows:  In Sec. \ref{sec:Hamiltonian} we shall introduce the Kondo Hamiltonian 
for the system. Its derivation is detailed in Appendix~\ref{app:SW}, by performing a  
Schrieffer-Wolff transformation~\cite{Schrieffer.66} to the single impurity Anderson model
that takes into account the coupling to the Majorana modes explicitly. 
In Sec. \ref{sec:AC_E} we shall focus on the equilibrium regime, for which we employ the NRG
technique to investigate the conductance and noise spectra in  both the 
 weak coupling  (Sec.~\ref{sec:WC}), and the
 strong coupling limit (Sec.~\ref{sec:SC}).
In Sec. \ref{sec:FRG} we construct the FRG  approach and use it later in Sec.~\ref{app:Green} to compute
the non-equilibrium conductance in the perturbative regime. 
We give our final remarks in Sec.~\ref{sec:concl}.

\section{Hamiltonian}\label{sec:Hamiltonian}
We shall assume that there is a single spin $S=1/2$
in the dot, that couples to the electrons in the normal lead and to the 
Majorana fermions through the Kondo interaction
\begin{equation}\label{eq:H_int}
H_{\rm int} = H_{\rm int}^{NM}+
\sum_{\s\s'} {J_{M}\over 2} \left \{\, 
\psi_{\s}^{\dagger}\,  \boldsymbol{\s}_{\s\s'} \Phi_{\s'}\, \mathbf{S}
+\textrm{H.c.} \right \}\, .
\end{equation}
The first term describes the regular Kondo interaction between the localized spin and the
normal lead, while the second 
term describes the coupling of the normal electrons to the Majorana fermions, mediated by 
the localized spin through the exchange interaction $J_M$. 
Such an "anomalous" term arises when performing the Schrieffer-Wolff transformations 
of an underlying Anderson model (Appendix \ref{app:SW}). Here, 
$\Phi_{\s}= u_\s\gamma_1 \sqrt{2}= u_\s (f+f^\dagger)$ describes the Majorana field.
As we shall see in Sec.~\ref{sec:WC}, 
the relevant quantity that enters the transport equations is the asymmetry of the dot defined as
$p = J_M/J$. Here we only address the strongly anisotropic situation corresponding to $p\ll 1$.
In the configuration presented in Fig.~\ref{fig:sketch},  the Majorana mode $\g_1$ 
is the one that couples to the dot. We chose it to be polarized along the 
spin $\downarrow_y$ direction, so it couples to the y-component of the 
dot spin~\cite{Leijnse.11}. Then, the couplings to spin-$\uparrow_z$ and spin-$\down_z$   directions
are the same in amplitude but have different phases $u_\s = 1/\sqrt{2}\{1, -i \}^{T}$.  In our convention
\be
\gamma_1 = \frac{1}{\sqrt{2}}(f+f^\dagger)\, , 
\label{eq:gamma_1}
\ee
while the other mode $\gamma_2= i(f-f^\dagger)/\sqrt{2}$ resides at the other end of the topological superconductor. 
In Eq.~\eqref{eq:gamma_1},  $f^{\dagger}$ are ordinary fermionic operators, which implies that the MBS operators $\gamma_n$ 
anticommute, $\{\gamma_1, \gamma_2\}=0$, but also satisfy the relation $\gamma_n^2 =1$. 
Notice that $f$ carries no spin index, so the occupation of the Majorana mode is 
$\langle n_f\rangle =\langle f^\dagger f\rangle =\{0,1\}$ only. 
The normal lead is described by the Hamiltonian 
\begin{equation}
H_N = \sum_{\s} \int_{-D}^{D} (\e-\mu_N)\, c_{\s}^\dagger(\e)\, c_{\s}(\e)\,d\e\, , 
\label{eq:H_N}
\end{equation}
with $\mu_N$ the chemical potential given by $\mu_N = eV/2$.
The electrons in the leads are assumed to be non-interacting, and 
the energy is measured from the chemical potential. The operators 
in Eq.~\eqref{eq:H_N}
satisfy the usual anti-commutation relations 
\begin{equation}
\{c_{\s}(\e), c_{\s'}^\dagger(\e')\} = \delta_{\s\s'}\delta(\e-\e').
\end{equation}
In the $f$-representation, the  
 Hamiltonian that describes the MBS is:
 \begin{equation}
H_M = (\xi-\mu_M) \,(f^\dagger\, f-1/2)\, ,
\label{eq:H_M}
\end{equation} 
with $\mu_M = -eV/2$. 
An external magnetic field enters through the dot Zeeman splitting, 
\begin{equation}
H_Z = \mathbf{B}\,\mathbf{S}\, , 
\label{H_Z}
\end{equation}
and is assumed to have no effect on the dynamics 
of the external normal lead~\footnote{Actually the magnetic field is renormalized by the coupling to the 
MBS (see Appendix \ref{app:SW}), as a similar term emerges from the Schrieffer-Wolff transformations.}. 
The total Hamiltonian 
\begin{equation}\label{eq:H_T}
H_T=H_N+H_M+H_{\rm int}\, , 
\end{equation}
captures the essential physics for our set-up. Under equilibrium 
conditions, $\mu_N=\mu_R=0$, the problem can be solved exactly~\cite{Wilson.75} 
by mapping the system to a Wilson chain and solving it with the help of NRG method~\cite{BudapestNRG}. We first
note that $H_T$ has a $Z_2$ symmetry, 
as it conserves the parity $P$ of the total fermionic number~\footnote{ 
When $\xi=0$,  the Hamiltonian has a SO(3) symmetry. 
A finite $\xi$ breaks this symmetry down to
$Z_2$, which corresponds to the parity operator.}. Because 
of the  anomalous terms~$\sim \psi_{\s}^\dagger f^\dagger$
that enter the interaction Hamiltonian~\eqref{eq:H_int}, neither the total charge 
nor the total spin are good quantum numbers, 
as their corresponding operators 
 do not commute with the  Hamiltonian~\eqref{eq:H_T}. Therefore, we can 
use only one quantum number $P=\{e,o\}$, to classify the multiplets of the Hamiltonian~\eqref{eq:H_T} as having
even or odd symmetry~\footnote{Using only a $Z_2$ symmetry makes the NRG calculations numerically expensive.}.

\section{Equilibrium ac-conductance and noise} 
\label{sec:AC_E}
To analyze the transport, we have to define
the current operator.  With the convention that the 
current describes the flow of the electrons from the normal lead into the dot, 
the corresponding operator is obtained from the equation of motion for the charge operator 
\begin{equation}
Q= e N= e\int_{-D}^{D} c^\dagger(\e)\, c(\e)\,d\e\, ,
\end{equation} 
in the normal lead as:
 
\begin{equation}
I= e\dot N= {i\,e\over \hbar}\, \big [ N,\, H_{\rm int}\big ]\,, 
\end{equation}
and it can be expressed in terms of the  $F_\s = (\mathbf{S}\,\boldsymbol{\s}\, \psi )_\s$, 
the so called composite fermion operator
\begin{equation}\label{eq:I}
I =\frac{i\,e }{ 2\,\hbar}J_{M}\sum_{\s}\left \{F^\dagger_\s \Phi_\s - \textrm{H.c.}\right \}\, . 
\end{equation}
The ac-conductance can be obtained by using the Kubo formalism as: 
\begin{equation}
G(\w) = \left \{ (I,I)_\w^{\rm ret} - (I,I)^{\rm ret}_{\w =0} \right \}/\w\, ,
\end{equation}
with $(I,I)^{\rm ret}_{\w}$ the Fourier transform of the retarded, 
time dependent correlator
\begin{equation}
(I,I)(t) = -i \theta(t)\langle [I(t), I(0)]\rangle\,.
\end{equation} 
It is useful to introduce a new local operator, 
the composite Majorana operator $C_\s = (F^\dagger_\s \Phi_\s - \textrm{H.c})$, in 
terms of which the real part of the  ac-conductance can be expressed as:
\begin{equation}
G(\w) = \left ({2\,e^2\over h} \right ){\pi^2\, J_{M}^2 \over 8\, \w}\sum_{\s \s'} \varrho_{C_\s C_{\s'}}(\w)\, ,
\label{eq:G_w}
\end{equation}
with $\varrho_{C_\s C_\s}(\w)$ the spectral representation of $C_{\s}$.  The dc-limit of the conductance can be obtained by taking the $\w\rightarrow 0$ limit
in Eq. \eqref{eq:G_w}, $G = \lim_{\w\rightarrow 0} G(\w)$.  At equilibrium, 
the spectrum of the symmetric noise, $S(t)$, 
is related to the real part of the 
ac-conductance through the fluctuation-dissipation theorem~\cite{Kubo.66}
\begin{equation}
S(\w) = -\w \coth\left ({\w\over 2\, T}\right)G(\w). 
\label{eq:FD}
\end{equation}
Since $C_\sigma$ are local operators, their spectrum can be computed exactly by using the 
NRG method~\cite{Bulla.08}. As we are interested in the current across the dot,
the ac-conductance $G(\omega)$ depends only on the dimensionless coupling $j_M$. We analyze separately the two different regimes that 
usually emerge in Kondo problems: $(i)$ the low energy, \emph{strongly correlated regime}, where 
the parameters $\{T, \omega\}\ll T_K$ are small compared to the Kondo scale, and $(ii)$ the perturbartive, 
or the \emph{weak coupling regime}, $\{ T, \omega\}\gg T_K$, where the problem can be described 
perturbatively in $j_M$. 

\subsection{T=0, weak coupling regime, $\omega \gg T_K$}
\label{sec:WC}

When the exchange coupling in Eq.~\eqref{eq:H_int} is treated perturbatively, one can 
evaluate the noise order by order in $j_M$. In the zeroth order, using definitions
\eqref{eq:noise} and \eqref{eq:I}, the ac-noise is computed as 
\be
S(\omega \gg \xi)\approx-\left (\frac{e^2}{h} \right)\frac{3\pi^2}{2}\, \varrho_0\, J_M^2,
\ee
and by the fluctuation dissipation theorem, Eq.~\eqref{eq:FD}, the perturbative results for the ac-conductance is then:
\be
G(\omega) \approx \left (\frac{e^2}{h} \right) \frac{3\pi^2}{2}\, \frac{\varrho_0\, J_M^2}{|\omega|}.
\ee
Summing the higher order contributions introduces logarithmic corrections to the couplings.  
To compensate for these terms, it turns out that to  is sufficient to replace the bare coupling with its renormalized value
$j_M \to p/\ln(|\w|/T_K)$, which gives for the conductance
\be
G(\omega) \approx \left (\frac{e^2}{h} \right) \frac{3\pi^2}{2}\, \frac{1}{\varrho_0\, |\omega|} \frac{p^2}{\ln^2(|\omega |/T_K)} .
\label{eq:G_pert}
\ee
As explained in Sec.~\ref{sec:Introduction}, this result is  in contrast with the usual $ 1/\ln^2(\omega/T_K) $ dependence that develops in the regular Kondo model in the same limit. Because of the extra
$\omega$ dependence at the denominator in Eq.~\eqref{eq:G_pert}, the
ac-conductance is expected to increase more steeply as $\omega$ decreases toward $T_K$. 
Except for a prefactor, 
the finite temperature behavior of the differential dc-conductance follows a similar trend, 
as shown in Eq.~\eqref{eq:G_pert_T}.

\subsection{Strong coupling regime, $\omega\ll T_K$}
\label{sec:SC}
As the frequency approaches the Kondo scale, the leading logarithmic correction 
in \eqref{eq:G_pert} diverges, signalling the breakdown of the perturbation 
theory. Symmetry considerations indicate that in the strongly correlated regime,   
there is a clear distinction between the limits when the MBS are 
hybridizing ($\xi\ne 0$) or not ($\xi=0$). 
When $\xi=0$, the multiplets with $\{e,o\}$
symmetry are degenerate in energy, 
and the transmission across the dot 
is at resonance. Below $T_K$, the transport is realized by inelastic
cotunneling processes by exciting parity degrees of freedom. 
In the $T=0$ limit and for $\xi=0$, by the Friedel sum rule, a half-unitary conductance is expected
\be
G(\{\omega, T\}=0) =2\;\Big (\frac{e^2}{h}\Big) \sin^2\Big({\pi\langle n_{\rm tot}\rangle \over 2}\Big)=\frac{e^2}{h}\, , 
\ee
as the average occupation of the QD-MBS system is $\langle n_{\rm tot}\rangle= 3/2$. 
In the small energy limit, 
$G(\omega) = G(\omega=0)+\dots$ remains practically constant~\footnote{
Fermi liquid theory guarantees that the conductance may display  quadratic $\sim \omega^2$ 
corrections.}.
As a result, the low 
frequency noise is linear in frequency
\be\label{eq:noise_linear}
S(\omega)=-\left(\frac{e^2}{h}\right )\,|\,\omega\,| .
\ee
These estimates are nicely captured by the NRG results presented in 
Fig.~\ref{fig:G_omega}. The linear $S (\omega)\sim \omega$ dependence of the noise, 
is also predicted in the regular Kondo problem~\cite{Sindel.05}, so that as long as the MBS are completely decoupled, 
$S(\omega)$ has the same linear dependence. 


When the degeneracy of the multiplets is broken ($\xi \ne 0$), there is a dramatic change in the conductance,
and a smaller energy scale $T^*\ll T_K$ given by
\begin{eqnarray}
T^*\simeq  \varrho_0 \Big ({\xi\over p}\Big)^2
\end{eqnarray}
emerges. At large enough frequencies, but still in the strong coupling regime, $T^*\ll\omega\ll T_K$, the cotunneling processes are still responsible for transport, and implicitly $G(\omega)\approx e^2/h$.
In the small frequency limit, $\omega\ll T^*$, the cotunneling processes are strongly suppressed. As the ground state becomes
non-degenerate in the  $\{e,o\}$ space, only higher order processes survive, in which at least two 
particles are exchanged with the normal lead. As a result, the ac-conductance 
is strongly suppressed below $T^*$,
\be
 G(\omega) \approx \alpha \left (\frac{e^2}{h} \right)\left (\frac{\omega}{T^*}\right )^2\;, 
\ee 
with $\alpha$ the same constant as in Eq.~\eqref{eq:noise_scaling}. In this regime, by 
fluctuation-dissipation theorem, the noise acquires a $S(\omega)\sim \omega^3$ dependence. 
Although the transport is governed by higher order processes, the NRG spectrum 
guarantees that the 
new fixed point is still that of a Fermi liquid in nature. 

\begin{figure}[!tbhp]
\begin{center}
\includegraphics[width= .95\columnwidth, clip]{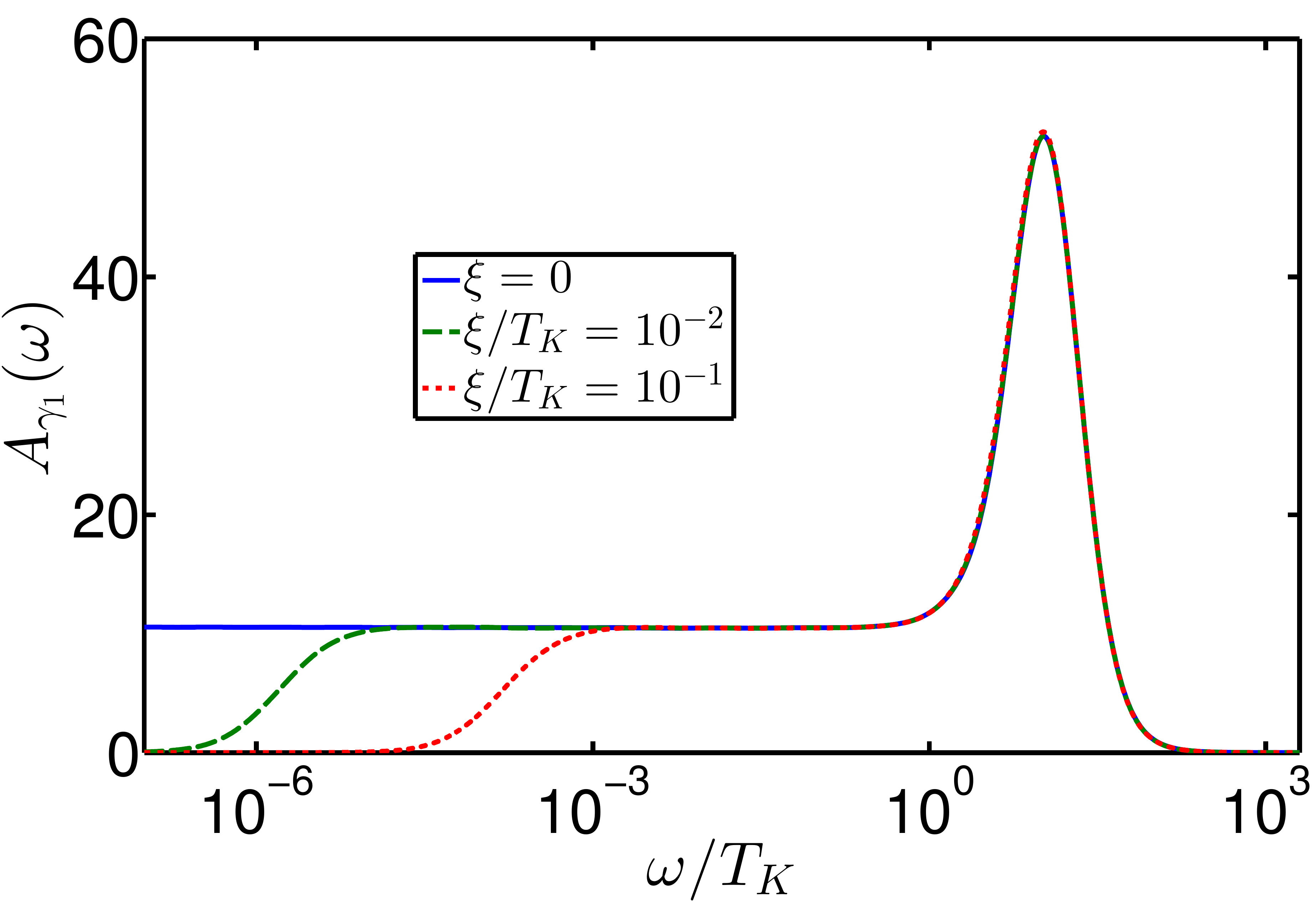}
\end{center}
\caption{\label{fig:gamma1_w}	
(Color online) Frequency dependent spectral function $A_{\gamma_1}(\omega)$ at $T=0$ for different values of $\xi$.}
\end{figure}

The characteristic energy $T^*$ affects  the spectral functions
of all the local operators. In Fig.~\ref{fig:gamma1_w} we represent the 
spectral function $A_{\g_1}(\w)$ of the Majorana operator $\g_1$ defined in Eq.~\eqref{eq:gamma_1}.  We can see  
that $A_{\g_1}(\w)$ shows a similar dependence on frequency to $G(\omega)$, and gets
suppressed below $T^*$. This can be understood by that fact that 
both the normal and the anomalous spectral functions of the f-operator used to construct
$A_{\g_1}(\w)$
display a large peak close to zero frequency with a width of the order of $T^*$. These large peaks in the two functions
have opposite signs, leading to a total suppression of $A_{\g_1}(\w)$ below $T^*$.
When $\xi\to 0$, each peak narrows down and increases in height, transforming into a zero-energy  $\delta$-peak,
which can be associated with the decoupled Majorana modes~\cite{Zitko.11}. 
\begin{figure}[tbhp]
\begin{center}
\includegraphics[width= .95\columnwidth, clip]{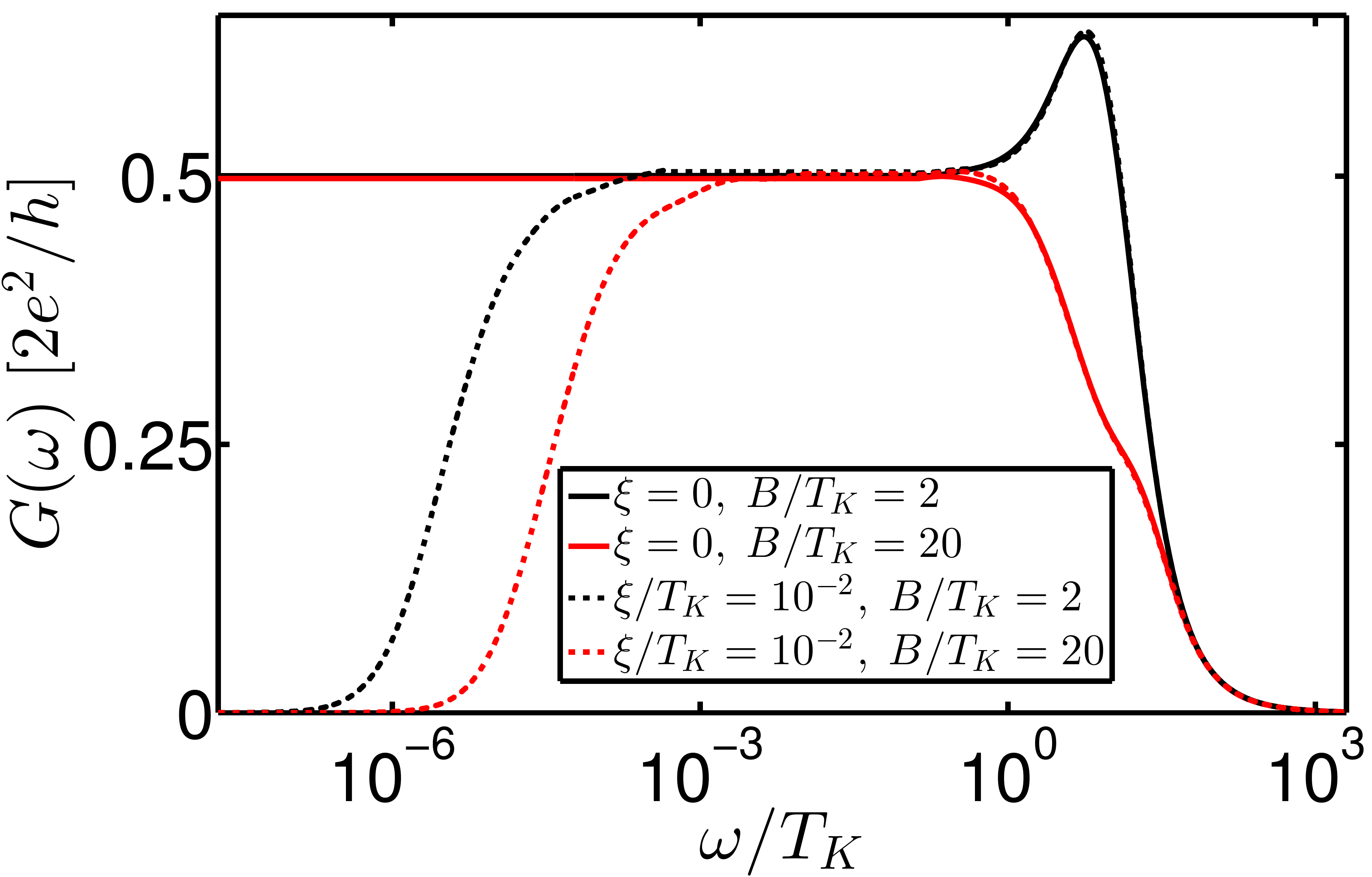}
\end{center}
\caption{(Color online) \label{fig:G_B}Magnetic field dependence of the ac-conductance for $\xi=0$ and $\xi/T_K=10^{-2}$, computed with the NRG.}
\end{figure}
The effect of a  non zero magnetic field is displayed in Fig.~\ref{fig:G_B}. As expected, a finite $B$ decreases the Kondo 
temperature, but as long as $\xi=0$, the low-frequency conductance remains unaffected. For finite $\xi$, we have found that
$T^*$ increases logarithmically with $B$ for $B\geq T_K$, but the conductance still displays the $\sim \omega^2$ behavior below $T^*$.
Although we apply an external magnetic field, we do not separate the conductance into spin channels, 
as the spin is not a good quantum number.

\section{FRG approach and the Keldysh action}\label{sec:FRG}

In contrast to the equilibrium situation, the non-equilibrium problem with $H_T$ defined in Sec.~\ref{sec:Hamiltonian} does
not possess a complete solution, not even numerically. 
That is because below the Kondo scale, $T_K$, defined in $\eqref{eq:T_K}$,
the couplings get strongly renormalized, and eventually diverge as the problem scales towards 
the strong coupling fixed point. 

In the present section, we shall construct a framework for 
computing the conductance under non-equilibrium conditions in the weak coupling regime,
under slightly more relaxed conditions,  where at least one of 
the energy scales involved is larger than $T_K$, i.e. $\max\{eV, T, B\}\gg  T_K$.
We shall discuss the major aspects of the method:  $(i)$ how the RG equations for the couplings can be obtained
and solved, and $(ii)$ analyze possible channels of decoherence of the local spin.

As we are dealing with a non-equilibrium situation, we use the  Keldysh formalism, and for that 
we first express the local spin ${\mathbf S}$ in terms of the 
pseudo-fermion operators $a_s$ by using the Abrikosov 
representation~\cite{Abrikosov.68}
\be
{\mathbf S} = {1\over 2}\sum_{ss'}a^{\dagger}_s\, {\boldsymbol \s_{ss'}}\,a_{s'}.
\ee  
They give the correct physical states provided that the dot is 
singly occupied. To ensure this constraint, we have to add a Lagrange multiplier 
to the Hamiltonian: $H_T\rightarrow H_T+\lambda\sum_sa^\dagger_s a_s$. 
We also switch to the path integral formalism.
The partition function is constructed in the usual way by replacing each fermionic field by a Grassmann 
variable~\cite{Negele}  
\be
\{\psi_{\s}, \psi^\dagger_{\s}, \g_i, \g_i^\dagger \}
\rightarrow 
\{\psi^{\kappa}_{\s}, \bar \psi^{\kappa}_{\s}, \g_i^{\kappa}, \bar \g_i^{\kappa} \}\, ,
\ee
  with
$\kappa =\{1,2\}$ labeling the upper/lower Keldysh contours.
These Grassmann fields depend now on time, and it is customary to 
write the path integral for the statistical sum as:
\begin{equation}
{\cal Z} =\int \left [{\cal D} \bar\psi\,{\cal D}\psi\, {\cal D} \bar \g \,{\cal D} \g {\cal D}\bar a \,{\cal D} a\right] 
e^{-{\cal S}[\bar \psi, \psi, \bar \g, \g, a, \bar a]}\,,
\end{equation}
with the total fermionic action: $
{\cal S} ={\cal S}_{ N}+S_{M}+{\cal S}_{\rm spin}+{\cal S}_{\rm int}.
$
The non-interaction part, ${\cal S}_{ N}$, describes the electrons in the 
normal lead: 
\be
{\cal S}_{N} =  \sum_{\kappa\kappa'}\sum_{\s}\int dt\, dt'
\bar\psi^{\kappa}_{\s}(t)\, \cG_{\s}^{\kappa\kappa' (0) -1} (t-t')\,
\psi^{\kappa'}_{\s}(t')\, .
\ee
Here $\cG_{\s}^{\kappa\kappa' (0)} (t)$ 
are the four components of the Keldysh 
Green's function (see Appendix~\ref{app:SF}). Notice that
$\cG_{\s}^{\kappa\kappa'(0)} (t)$ is 
diagonal in the  spin labels, but has off-diagonal 
elements in the Keldysh indices. For the MBS, the non-interacting action 
is
\be
{\cal S}_{M} =  \sum_{\kappa\kappa'}\sum_{i,j=\{1,2\}}\int dt\, dt'
\bar\g^{\kappa}_{i}(t)\, \Xi_{ij}^{\kappa\kappa' (0) -1} (t-t')\,
\g^{\kappa'}_{j}(t'),
\ee
where $\Xi_{ij}^{\kappa\kappa' (0)}$ are again the four components 
of the corresponding Keldysh 
Green's function (see Appendix~\ref{app:SF}), 
and we have a similar ${\cal S}_{\rm spin}$ term for the localized spin.
These kinetic terms $S_{M}$, $S_{N}$ and $S_{\rm spin}$ are quadratic in the 
fields and determine the non-interaction Green's function for each species of 
fermions. 
The interaction part, ${\cal S}_{\rm int} \propto \int H_{\rm int}dt $, can be written 
in a relatively general form by considering that the
exchange interaction  is no longer 
an instantaneous process, but acquires a time dependence~\cite{Moca.14}
\begin{eqnarray}
 j\, {\boldsymbol\s}_{\s\s'} \otimes{\boldsymbol\s}_{ss'}\delta(t)
 &= &j_{ s s'}^{\s\s'}\,\delta(t)
  \rightarrow  
g^{\s \s'}_{ss'}(t)\nonumber\\
 j_M\, {\boldsymbol\s}_{\s\s'} \otimes{\boldsymbol\s}_{ss'}\delta(t)
 &= &v_{ s s'}^{\s\s'}\,\delta(t)
  \rightarrow  
v^{\s \s'}_{ss'}(t)\, . \label{eq:couplings_def}
\end{eqnarray}
Replacing the couplings by the time dependent ones, the interacting action becomes
\begin{multline}
{\cal S}_{\rm int} =  \sum_{\kappa}\frac{(-i)^{\kappa}}{4}\sum_{\s\s'}\int dt\, dt'
\left\{
g_{s s'}^{\s \s'}(t-t')\right .\times \\
\bar \psi^{\kappa}_{\s}(t)\, \psi^{\kappa}_{\s'}(t')\,
\bar a^{\kappa}_{\s}(T)\, a^{\kappa}_{\s'}(T)+\\
\left . \left ( v_{s s'}^{\s \s'}(t-t')\,
\bar \psi^{\kappa}_{\s}(t)\, \Phi^{\kappa}_{\s'}(t')\,+H.c.\right )
\bar a^{\kappa}_{\s}(T)\, a^{\kappa}_{\s'}(T) \right . \, . 
\label{eq:S_int}
\end{multline}
The first term in  ${\cal S_{\rm int}}$  describes processes 
where the electrons are scattered in the normal lead, and are responsible for the 
regular Kondo effect. 
The second term involves the Majorana modes, 
and describes processes in which one electron at time $t$ is scattered by 
the spin fluctuation into the MBS at time $t'$.
The spin fluctuation itself happens at some intermediate time $T= (t+t')/2$. 
The electronic processes are considered to be 
instantaneous, without any decay.  We  neglect the 
self-energy of the conduction electrons in the leads, but in contrast, the much slower
spin flip processes are characterized by a decay rate that will be 
discussed in more detail in Sec.~\ref{sec:Decoherence}.

\subsection{RG equations for the dimensionless couplings}
\label{sec:RG}

\begin{figure}[!htb]
\begin{center}
\includegraphics[width= .95\columnwidth, clip]{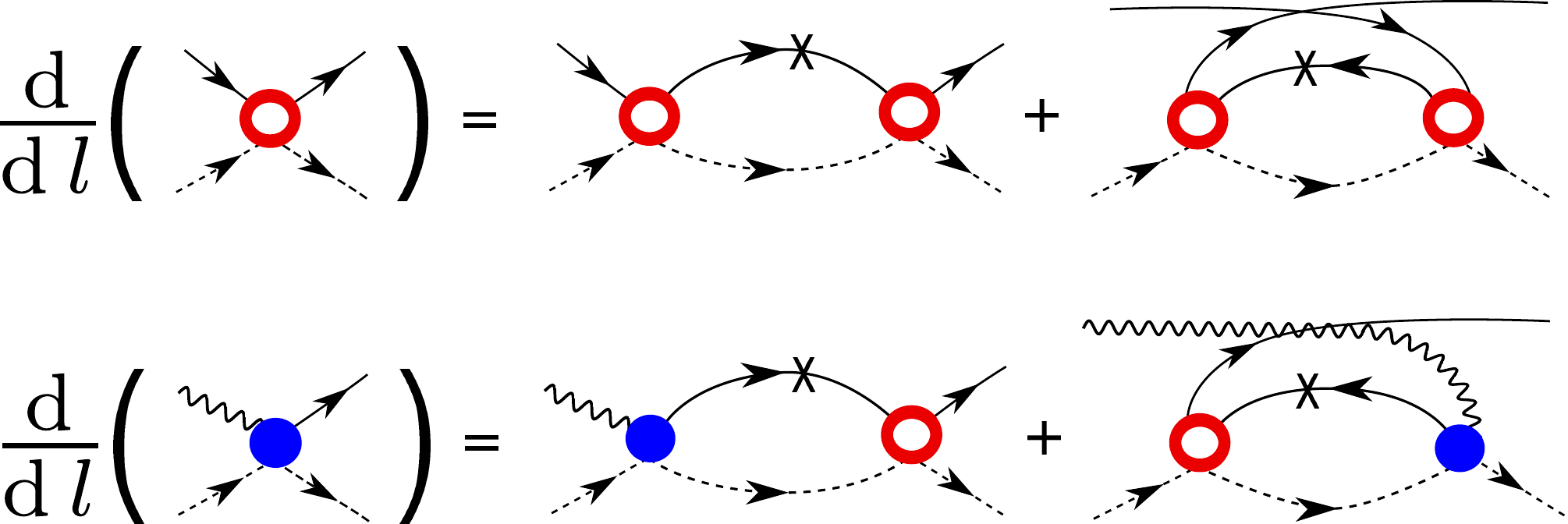}
\end{center}
\caption{
(a) Graphical representation of the differential equation for the couplings.
Blue (filled) dots denote the coupling $v_{ss'}^{\s\s'}(\omega)$,  while red (hollow) dots denote $g_{ss'}^{\s\s'}(\omega)$.
The derivatives of the electronic Green's functions with respect to $\ln D$ are represented by crossed solid lines. 
The pseudofermions are represented by dashed, and the Majorana Green's functions by 
wavy lines. 
}\label{fig:couplings}	
\end{figure}

\begin{figure}[!tbhp]
\begin{center}
\includegraphics[width= 0.9\columnwidth]{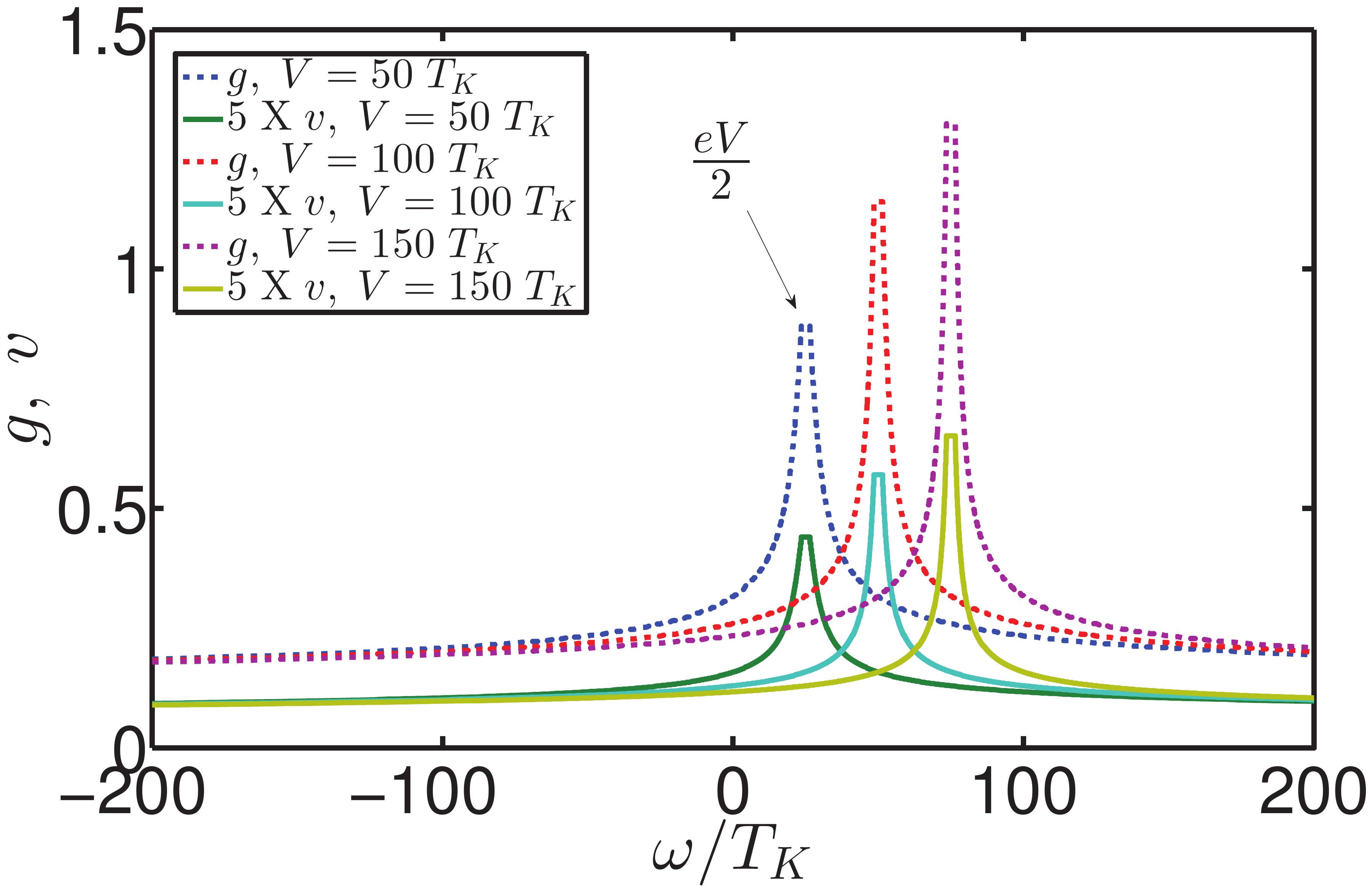}
\end{center}	
\caption{\label{fig:sc_eq_B_0}
(Color online) Renormalized couplings for three different voltages at zero 
magnetic field, taking $p=0.1$.  Both $g(\w)$ and 
$v(\w)$ show 
a logarithmic singularity at $\omega =eV/2$.  }
\end{figure}

In this section we shall construct the scaling equations for 
the exchange couplings by following the work in Ref.~\cite{Moca.14}.
The general recipe consists of expanding the action ${\cal S}$
in terms of ${\cal S}_{\rm int}$ given in Eq.~\eqref{eq:S_int}, followed by 
a rescaling of the cutoff parameter $a\rightarrow a'$. 
It leads to a set of integro-differential equations for the renormalized couplings. 
As noticed in Ref.~\cite{Moca.14}, it is simpler to analyze these equations in the Fourier space, as 
they acquire a simpler structure. We shall not present the full derivation here, but only give their final form
%
\begin{eqnarray}
\frac{ \rmd\; g_{ss'}^{\s \s'} (\omega) }{\rmd l}  &=& 
\frac{1}{4} \sum_{\tilde \sigma \tilde s}
\left [ g_{\tilde s s'}^{\sigma  \tilde \sigma } (\omega_{\tilde s s}) 
\;g_{s \tilde s}^{ \tilde \sigma \sigma' } (\omega_{\tilde s s'} )
\Theta_{{\omega_{\tilde s s} + \omega_{\tilde s s'} \over 2} - \mu_{N}}
\right .  \nonumber\\
& & -\left . g_{s \tilde s}^{ \sigma  \tilde \sigma} (\omega_{s'\tilde s}) \right .   \;g_{\tilde s s'}^{\tilde \sigma  \sigma'} 
(\omega_{s\tilde s}) 
\left . \Theta_{{
 \omega_{s\tilde s} +\omega_{s'\tilde s}) \over 2}  -\mu_{N}  }
\right ]\label{eq:scaling_g}\nonumber\\
\frac{ \rmd\; v_{ss'}^{\s \s'} (\omega) }{\rmd l}  &=& 
\frac{1}{4} \sum_{\tilde \alpha\tilde \sigma \tilde s}
\left [ g_{\tilde s s'}^{\sigma  \tilde \sigma } (\omega_{\tilde s s}) 
\;v_{s\tilde s}^{ \tilde \sigma \sigma' } (\omega_{\tilde s s'} )
\Theta_{{\omega_{\tilde s s} + \omega_{\tilde s s'} \over 2} - \mu_{N}}
\right . \nonumber\\
& & -\left . g_{s \tilde s}^{ \sigma  \tilde \sigma} (\omega_{s'\tilde s}) \right .    \;v_{\tilde s s'}^{\tilde \sigma  \sigma'} 
(\omega_{s\tilde s}) 
\left . \Theta_{{
 \omega_{s\tilde s} +\omega_{s'\tilde s} \over 2}  -\mu_{N}  }
\right ] \,.\nonumber\\
\end{eqnarray}
 %
Here $l = \ln(a/a_0)$ is the scaling variable, and $a_0$ is the initial value 
of the cut-off time. We also used the notations $\omega_{ss'} =\omega+(\lambda_s-\lambda_{s'})/2$, while 
$\Theta_{  \omega }=\Theta(1/a -|\omega  + i\, \G|)$
is some cutoff function that can be chosen to be the Heaviside step function~\cite{Rosch.05}, 
with  $\G$  the  relaxation rate serving as a cut-off energy (see Sec. \ref{sec:Decoherence}). 
The pseudo-fermion energies are simply given by $\lambda_s = -s B/2$.
To get a better understanding of the processes involved, we illustrate in Fig.~\ref{fig:couplings}  
a diagrammatic representation for Eqs.~\eqref{eq:scaling_g}. 
The solid blue vertex is associated with $j_M$, while the wavy lines are the Majorana propagators. 
%
In the absence of an 
external magnetic field, $\lambda_s =0$, and by using the symmetries of the 
coupling matrix, it is possible to reduce \eqref{eq:scaling_g}
 to a much simpler form~\cite{Moca.14}
\begin{eqnarray}
\frac{ \rmd\; { g}(\omega) }{\rmd l} & 
= &{ g}(\omega)\, {\Theta(\omega, a)}\,{g}(\omega)\label{eq:rg_B_0} \\
\frac{ \rmd\; {v}(\omega) }{\rmd l} &
= &{g}(\omega)\, { \Theta(\omega, a)}\,{ v}(\omega) \, . \label{eq:rg_B_1}
\end{eqnarray}
Here ${\Theta(\omega, a)} =\Theta_{\omega-\mu_N} $. 
 The equation for $g(\omega)$ is independent of $v(\omega)$,
and solely defines the Kondo scale $T_K$. To understand that, we notice that 
in the static limit,  $\omega\to 0$, Eq.~\eqref{eq:rg_B_0} 
reduces to the 'poor man's equation', as first derived by Anderson~\cite{Anderson.70},
with the asymptotic solution $g\approx 1/\ln(D/T_K)$, and $T_K$ defined in Eq. (\ref{eq:T_K}).
The other equation can also be solved analytically in the same limit, and a similar behavior is found,
 namely $v\approx p/\ln(D/T_K)$. On the other hand, in general,
 the coupled set of equations \eqref{eq:rg_B_0}-\eqref{eq:rg_B_1} can be solved only numerically.
Typical results for the renormalized couplings in the absence of the 
external magnetic field, are presented in 
 Fig.~\ref{fig:sc_eq_B_0}.   The couplings show a single logarithmic divergence
at frequency $\omega \simeq  eV/2$, which corresponds to the voltage 
applied to the normal lead.  In the very high-frequency limit, the couplings lose their logarithmic behavior 
and tend to a constant value corresponding to the bare couplings.

\begin{figure}[!tbhp]
\begin{center}
\includegraphics[width= 0.9\columnwidth]{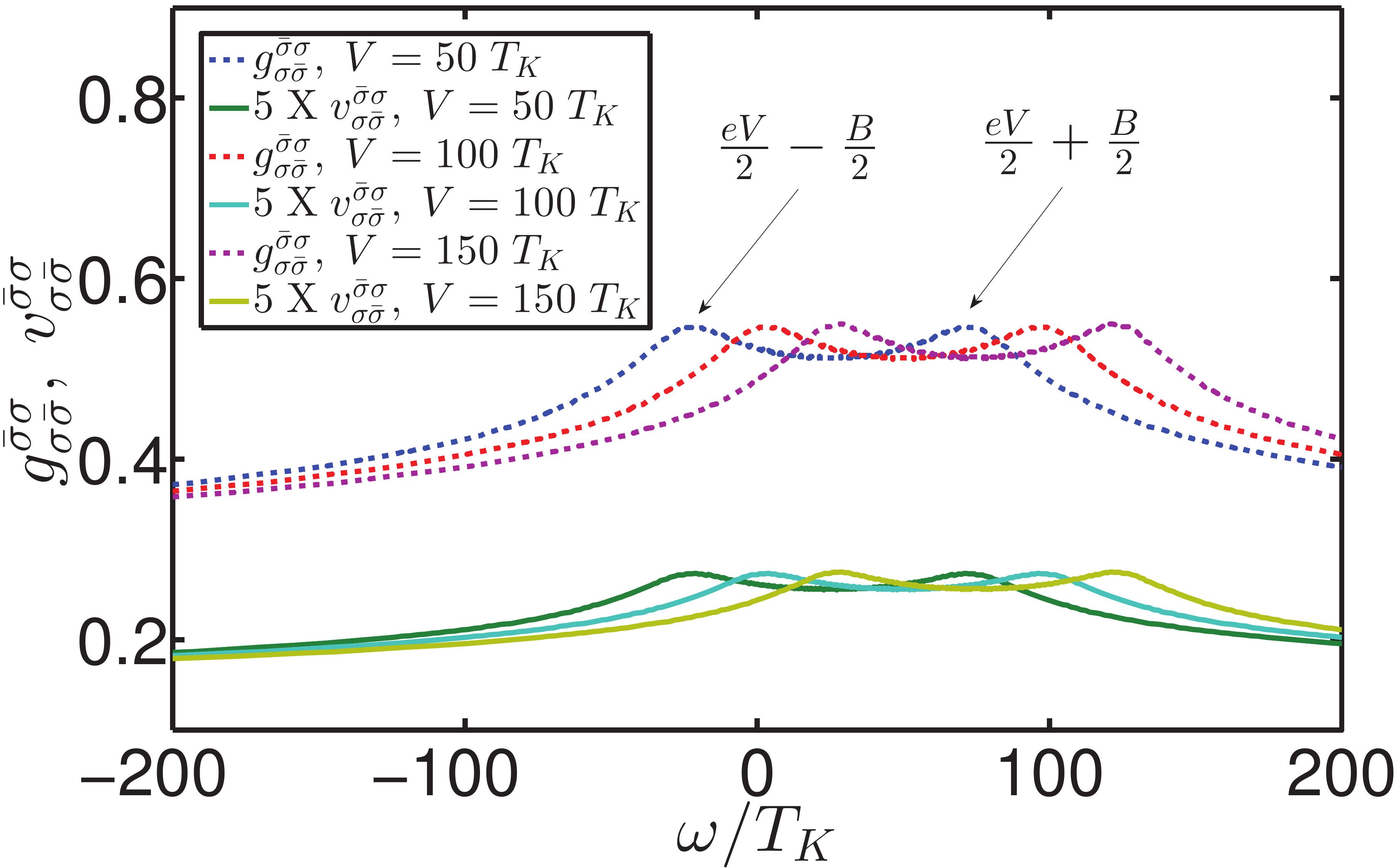}
\end{center}
\caption{\label{fig:sc_eq_B}
(Color online) Some of the components of the dimensionless renormalized couplings $g_{s s'}^{ \s  \s'}(\omega)$ and 
$v_{s s'}^{ \s  \s'}(\omega)$ in the presence of an external magnetic field $B=100\, T_K$. In the present calculation we  fixed $p=0.1$. The logarithmic singularity
at $eV/2$ when B=0 is now broadened and split into two separate peaks at $(eV\pm B)/2$.}
\label{fig:sc_eq_B_finite}	
\end{figure}
Applying an external magnetic field, the logarithmic divergence at $\omega \simeq eV/2$
is split, and the couplings develop peaks at $\omega \simeq (eV\pm B)/2$.
In  Fig.~\ref{fig:sc_eq_B_finite} we present the results for some of the components of 
$\mathbf g(\omega)\equiv g_{s  s'}^{ \sigma \sigma'}$ and $\mathbf v(\omega)\equiv v_{s  s'}^{ \sigma \sigma'}$, when a magnetic field of $B=100\, T_K$ is applied. 
As the relaxation rate increases with $B$, the peaks are broader and reduced in amplitude as compared to the $B=0$ case.

\subsection{Decoherence effects}
\label{sec:Decoherence}

In this section we shall use the master equation approach (ME)  
to give a quantitative description of the relaxation processes. 
Within the ME approach we usually start by considering that initially
the dot is in a certain spin configuration: either in a 
spin-up ($S=\uparrow$) or spin-down ($S=\downarrow$) state, then  
let the system evolve from one state to another by means of scattering processes
between the NM and the MBS leads.   
As these processes are random, they are characterized 
by some scattering rates, $\gamma^{S'\leftarrow S}_{F \leftarrow I}$. These rates
 describe
the cotunneling processes in which an electron leaves the "lead" $
\alpha=\{NM,MBS\}$ in state $I$, while another electron
 enters "lead" $\beta=\{NM,MBS\}$ in state $F$, and at the same time 
the localized spin is flipped, going 
from $S\to S'$. At finite temperatures, the scattering into the
NM lead alone dominates. These processes take a random time of the order of $\tau_\G \approx  h/(\pi j^2 T)$, as we shall see below. 
Therefore, it is justified to introduce a typical time scale
$\tau_\G\approx h/\G$ and a typical energy $\G$ that characterizes the 
relaxation processes. 
On the contrary, at $T=0$ and $B=0$, the only relevant scattering processes 
are those where the electrons/holes are exchanged between the NM and MBS.
Processes where only the Majorana lead is involved, are  $\propto |\eta|^4$ 
and will be neglected. 
Here $\eta$ is the hopping matrix between the QD and 
the Majorana lead. 
Considering for example the processes in which an electron is removed from the normal lead and one electron is added to the MBS (in either order), 
we get the rate
\begin{eqnarray}
\gamma^{\Uparrow \leftarrow \Downarrow}_{F_a \leftarrow I_a} &=&\frac{\pi}{4\varrho_0}\int d\epsilon\, f(\epsilon-\mu_L)\left[1-f(\epsilon-\mu_R)\right]\nonumber\\
&&\left |j_{M} \right |^2 \delta(\epsilon-\xi-\mu_R)\, ,
\end{eqnarray}
where the index $a$ denotes these particular processes.
Then, the decay of the local spin is characterized by the rate:
\begin{eqnarray}\label{eq:rate}
\Gamma^{\Uparrow \leftarrow \Downarrow}&=&\Gamma^{\Downarrow \leftarrow \Uparrow}=\sum_{I,F}\gamma^{\Uparrow \leftarrow \Downarrow}_{F \leftarrow I} \\
&\simeq&\frac{\pi}{4\varrho_0} j_{M}^2\left[f(-\xi)f(\xi-eV)+f(\xi)f(-\xi+eV) \right]\, , \nonumber
\end{eqnarray}
and the  expectation value of the z-projection of the impurity spin satisfies the equation:
\begin{eqnarray}
\frac{d}{dt}\langle S_z \rangle=-(\Gamma^{\Uparrow \leftarrow \Downarrow}+\Gamma^{\Downarrow \leftarrow \Uparrow})\langle S_z\rangle.
\end{eqnarray}
Considering the case $\xi=0$ only, the total decoherence is
\begin{eqnarray}
\Gamma(T=0, V)&=& \frac{\pi j_{M}^2}{4\varrho_0}. 
\label{eq:Gamma_T0}
\end{eqnarray} 
At finite temperatures, the coupling to the normal lead dominates, but in  
general, the two mechanisms for the decoherence produce the total decay rate
\begin{equation}
\G \simeq \pi j^2 T +\frac{\pi j_M^2}{4 \varrho_0}.
\label{eq:Gamma}
\end{equation}
At first sight, the spin relaxation rate appears to be voltage independent, 
and only to depend on temperature. 
This is the simple minded estimate for $\Gamma$, as obtained
in the lowest order perturbation theory. 
Higher order corrections can be summed up simply 
by replacing in Eq.~\eqref{eq:Gamma} the bare couplings $j$ and $j_M$ by their renormalized value
\begin{equation}
{j \choose j_M}\rightarrow {1\choose p}\frac{1}{\ln(\max(|eV|,T)/T_K)}\, . 
\label{eq:RPT}
\end{equation}
In this way the 
relaxation rate acquires a weak, logarithmic voltage dependence.

\begin{figure}[!tbhp]
\begin{center}
\includegraphics[width=0.9\columnwidth]{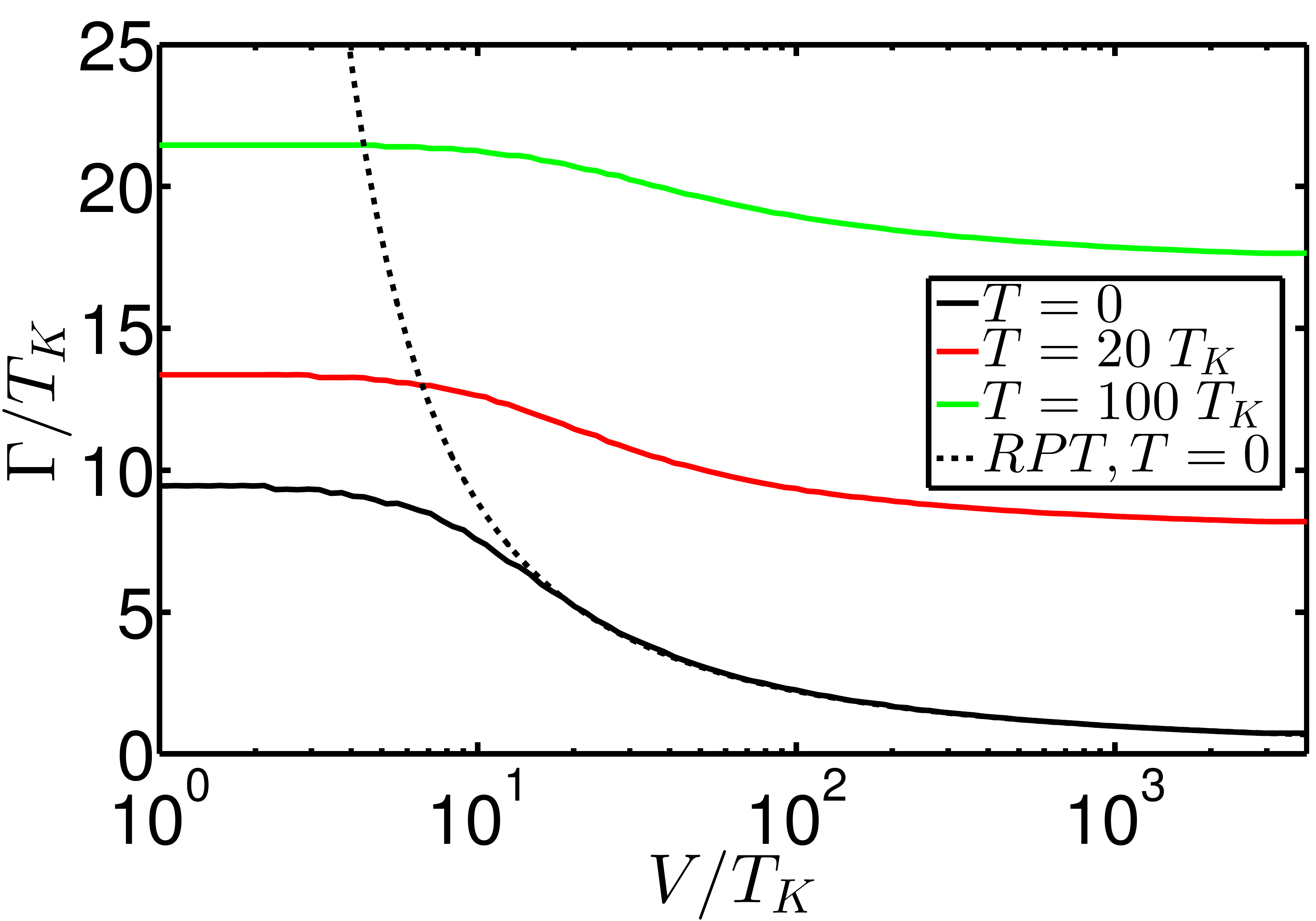}
\caption{(Color online) The decoherence as function of V for different temperatures. The dotted
line (RPT) is the renormalized perturbative result, \eqref{eq:Gamma_T0}, with the renormalized  coupling $j_M\to p/\ln(V/T_K)$}
\label{fig:Gamma_V}
\end{center}
\end{figure}

\begin{figure}[!tbhp]
\begin{center}
\includegraphics[width=0.9\columnwidth]{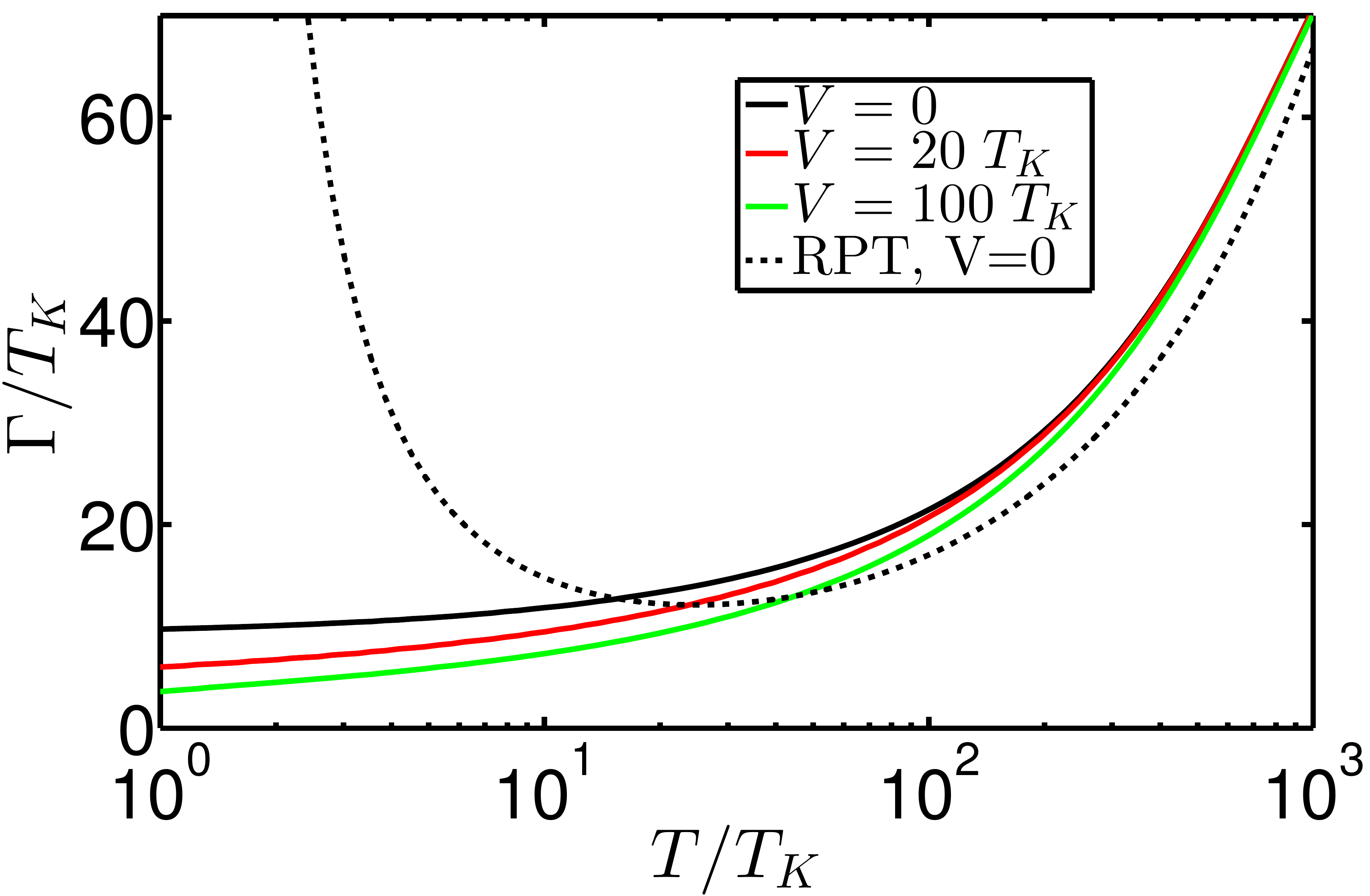}
\caption{(Color online) The decoherence $\Gamma$ as function of temperature. 
In the high temperature limit, $T\gg T_K$, the source of decoherence are
 the processes involving the normal lead. The black  dotted line 
 represents the renormalized perturbation theory (RPT) result, given 
 by \eqref{eq:Gamma}, but using the renormalized couplings.}
\label{fig:Gamma_T}
\end{center}
\end{figure}

To check our estimates, 
we have compared them with the more elaborate results obtained within the fRG approach,
where the full frequency renormalized couplings $\mathbf g(\omega)$ and $\mathbf v(\omega)$
were used to estimate $\Gamma$~\cite{Moca.14}.
Figs.~\ref{fig:Gamma_V} and~\ref{fig:Gamma_T} present such comparisons. 

The voltage dependence of $\Gamma$ is shown in Fig.~\ref{fig:Gamma_V}
for different temperatures. At $T=0$, the only available relaxation processes
involves resonant scattering between the  normal lead and the MBS. 
In that regard,
the high voltage limit, $eV\gg T_K$, 
is very well captured by our perturbative result, Eq.~\eqref{eq:Gamma_T0}.
On the other hand, the offsets at $T>0$ are due to the processes involving the normal lead. At any temperature, 
$\G(V)$ decreases with the
voltage, in contrast to the usual almost linear increase in the regular Kondo problem~\cite{Moca.14}.
The temperature dependence is presented in Fig.~\ref{fig:Gamma_T}. As the 
scattering processes involve now mostly the normal lead, the $\propto T/\ln^2 (T/T_K)$ dependence
is recaptured.

\section{Non-equilibrium differential conductance}
\label{app:Green}

In the present section we present results for the non-equilibrium differential conductance. 
We start by introducing the current operator~\cite{Moca.14}, which in terms of time dependent couplings, 
becomes a nonlocal operator
\begin{eqnarray}
I(t)&=&i \frac{ e}{2\hbar } \sum_{\s \s' s s'} \int {\rm dt'}v_{s s'}^{\s \s'}(t-t')\psi_{\s}^{\dagger}(t)u_{\s'}^{*}\left [f(t')+f^{\dagger}(t')\right]\nonumber\\
&&a_s^{\dagger}\left(\frac{t+t'}{2}\right)a_{s'}\left(\frac{t+t'}{2}\right)dt'+\mathrm{H.c.}\;.
\end{eqnarray}
Within 
the  perturbation theory on the Keldysh contour, it can be evaluated order by order in dimensionless couplings. 
Here we restrict ourselves to the first order in the perturbative expansion:
\begin{eqnarray}
\langle I (t) \rangle=(-i)  \int_K \langle {\cal T}_C I (t)\,  H_{\rm int}(t')\rangle \,{\rm d} t',
\label{eq:av_I}
\end{eqnarray}
with the time integral performed on the Keldysh contour. 
 We assume the symmetric bias configuration, where the finite voltages $\pm V/2$ are applied to the leads.
 When evaluating~\eqref{eq:av_I}, it is simpler to transform it to Fourier space, and compute the frequency integrals 
 numerically using for the renormalized couplings the solutions obtained in Eqs.~\eqref{eq:rg_B_0}. Then, the usual
 differential conductance is computed as
 \begin{equation}
 G(V, T) = \frac{d \langle I\rangle}{dV}\, .
 \end{equation}
 The results for $G(V)$ were already discussed to some extent 
 in Sec~\ref{sec:Introduction},  and indicate
 that in some limits, the differential conductance becomes negative. In this section we shall 
 extend our analysis, and try to present analytical results where possible.

\begin{figure}[!ptbh]
\begin{center}
\includegraphics[width= 0.9\columnwidth]{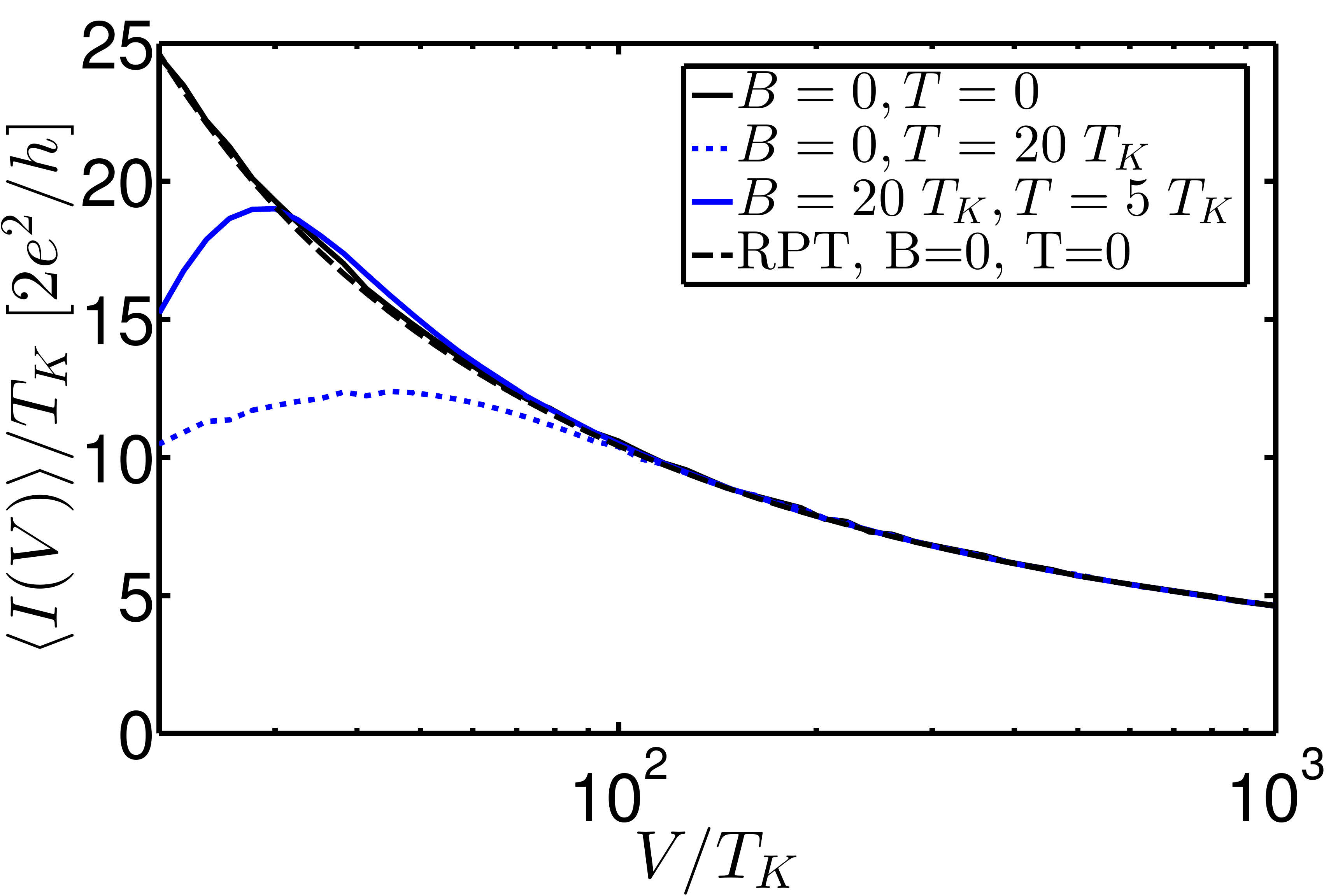}
\end{center}
\caption{(Color online)
The  average current $\langle I\rangle$ as function of the voltage drop V, for different
temperatures and magnetic fields. The blue dotted line is the renormalized perturbative result.
}\label{fig:I_V} 
\end{figure}

To obtain $G(V)$ at non-equilibrium, we have to calculate the average current, $\langle I\rangle $, 
to second order in $j_M$. 
 Evaluating Eq.~\eqref{eq:av_I} by using the Green's 
functions defined in Appendix.~\ref{app:SF}, 
 we obtain 
\begin{eqnarray}\label{eq:noneq_curr}
\langle I (V,T) \rangle&=&\frac{e}{h}\frac{3  \pi^2}{2  \varrho_0} j_{M}^2\left(-\frac{eV}{2}+\xi\right)\left[f(-\xi)f(\xi-eV)-\right.\nonumber\\
&& \left. - f(\xi)f(eV -\xi)\right].
\end{eqnarray}

This result can  be understood in terms of the scattering processes
depicted in Fig.~\ref{fig:diff_cond} (upper panel). The first term in the bracket can be associated with a process in which 
an electron is scattered from a filled state in the normal lead into an available MBS state, while the second term is 
the backward process. In a given voltage configuration, because of the Fermi functions that enter
Eq.~\eqref{eq:noneq_curr}, at $T=0$ and finite $\xi$,
the device acts as a diode that conducts only in one direction. Evaluating higher order 
contributions to the current, one obtains logarithmic corrections to the coupling $j_M$. This is 
depicted in the upper panel of Fig.~\ref{fig:diff_cond}, which indicates that the logarithmic corrections in the renormalized 
perturbative procedure amounts to the substitution \eqref{eq:RPT}.
%
%
Then, in the high voltage limit, $|eV|\gg T_K$, and for $\{T,\xi\}\ll V$, the differential conductance has a somewhat simple analytical expression
\begin{eqnarray}\label{eq:GV}
&&G(V\gg T_K, T\ll V)\simeq\nonumber\\
&&\simeq\frac {e^2}{h}\frac{3 \pi^2}{2 \varrho_0}\frac{p^2}{\ln^2 \left ( \frac{|eV|}{T_K} \right)}\left [\;\frac{1}{4 T}\;\mathrm{sech}^2\left(\frac{-eV+\xi}{2T}\right ) - \right.
\nonumber \\
&&\left .- 2 \;\frac{f(-\xi)-f(-\xi+eV)}{\ln \left ( \frac{|eV|}{T_K} \right) eV} \right ]  .
\label{eq:diff_cond}
\end{eqnarray}
The first term in the bracket is the leading logarithmic correction which was first obtained in Ref.~\cite{Golub.11}. As 
expected, at finite $T$, this term dominates at intermediate voltages, but under some specific conditions, 
when for example $|eV|\gg T$, the subleading correction (second term in Eq.~\eqref{eq:diff_cond}) becomes the dominant one. The sign
of the latter term is negative due to the fact that $j_M(-eV/2)$ decreases with increasing the voltage.
Next, we will consider the case $\xi=0$ for simplicity.
At $T=0$, the result for $G(V)$ obtained from Eq.~\eqref{eq:diff_cond} is plotted together with the fRG result in 
Fig.~\ref{fig:diff_cond}, while the corresponding current curves are shown in Fig.~\ref{fig:I_V}. 
The renormalized perturbative result is in perfect agreement with the fRG result over a wide range of voltages as long as 
$eV\gg \mathrm{max}\{T_K,\Gamma\}$. At finite temperature, the leading term dominates at intermediate voltages, leading to a 
change of the sign in $G(V)$.  

\section{Conclusions}\label{sec:concl}
In this work we have investigated in detail the ac-conductance and noise through 
a quantum dot in the local moment regime, 
which is coupled to a normal metal lead on one side, and to a topological superconductor on the other side. 
Starting with a single impurity Anderson Hamiltonian, we have used the 
Schrieffer-Wolff transformations~\cite{Schrieffer.66} to map the problem  
to an effective Kondo model in  which the coupling of the 
local spin to the Majorana modes is explicit. 

The problem at equilibrium was investigated by using the powerful numerical renormalization group approach,
supplemented by the perturbative calculations in the weak coupling regime. 
The ac-conductance and noise has a spectacular behavior: When the Majorana modes are
coupled, a new energy scale $T^*$ emerges, that controls the low energy transport. 
We have identified this scale as a manifestation of the parity conservation in the absence of 
parity relaxation mechanisms. For frequencies below  $T^*$, the ac-conductance becomes suppressed, and  as a result,
the noise spectrum acquires a $S(\omega)\sim \omega ^3$ dependence. 
On the other hand, when the  
Majorana modes are not coupled, the low energy transport is entirely controlled by the Kondo physics, 
and transport is realized through cotunneling processes that excite the parity degrees of freedom.

To address the non-equilibrium situation, we developed a real time functional 
renormalization group (fRG) formalism and constructed a set of RG equations for the
renormalized couplings.
The structure of the RG equations is such that it sums up all the leading logarithmic 
contributions and is valid at any voltage $V$, temperature $T$, or frequency $\omega$,
under the condition that $\max\{eV, T, \omega\}\gg T_K$. This set of RG equations also indicates 
that the Kondo scale is not affected by the coupling to the Majorana modes, 
and is entirely controlled by the exchange with the normal lead.  
As the dynamics of 
the local spin is strongly affected by the decoherence effects, we have used 
a master equation approach to investigate 
how the spin decays, and we have computed  
the spin relaxation rate $\Gamma$.  We have found that $\Gamma$ has a slow logarithmic 
voltage dependence, but a faster $T/\ln^2(T/T_K)$ variation with temperature.  
The RG equations have been solved  
numerically in the Fourier space, and then their solutions have been used to compute the 
temperature and voltage dependence of the differential 
conductance. We have found that in the limit $T\to 0$,
the differential conductance becomes negative. 
Increasing the temperature can change the sign of the differential conductance. 
In the weak coupling regime, 
our results agree with the ones presented in Ref.~\cite{Golub.11}, and show that 
the logarithmic increase in the transport coefficients is altered,  and a much steeper increase emerges 
 as the energy drops towards the Kondo scale. Measuring 
 such anomalies in transport indicates the presence of the Majorana modes in the system. 

\section*{ Acknowledgments}
This research has been
supported by the Romanian UEFISCDI under French-Romanian Grant
DYMESYS (ANR 2011-IS04-001-01 and Contract
No. PN-II-ID-JRP-2011-1) and by 
the Hungarian Research Funds under grant Nos. 
K105149, CNK80991, TAMOP-4.2.1/B-09/1/KMR-2010-0002.

\appendix

\section{The Schrieffer Wolff transformations}\label{app:SW}

In our approach, the quantum dot is described by the single impurity Anderson
model (SIAM), which in its simplest form consists of an energy level of energy $\varepsilon_d$, and an 
on-site repulsion term
\begin{equation}\label{eq:H_Anderson}
H_{D}=\sum_{\sigma}\varepsilon_{d}d_{\sigma}^{\dagger}d_{\sigma}+U\, n_{\uparrow}n_{\downarrow},
\end{equation}
and can be coupled to several leads. In Eq.~\eqref{eq:H_Anderson}, $U$ represents the on-site Coulomb interaction. 
Here we consider the setup sketched in Fig.~\ref{fig:sketch},
in which the dot is coupled to the Fermi see of the conduction electrons of a metallic lead, 
described by the Hamiltonian~\eqref{eq:H_N},
and to a nanowire with induced superconductivity, characterized by an induced gap $\Delta$ and supporting Majorana 
bound states at its ends. As discussed in Sec. \ref{sec:Introduction}, in the small energy limit, $E\ll \Delta$, 
the topological superconductor is simply modeled by a pair of MBS states, described by the Hamiltonian~\eqref{eq:H_M}.
In its most general form, the hybridization between the dot and the two external leads is given 
by the tunneling Hamiltonian 
\begin{eqnarray}
H_{tun}&=&\sum_{\textbf{k},\sigma}(t_{\textbf{k}}c_{\textbf{k}\sigma}^{\dagger}d_{\sigma}
+t_{\textbf{k}}^{*}d_{\sigma}^{\dagger}c_{\textbf{k}\sigma})\nonumber \\
 &+&\sum_{\sigma}\eta_{\sigma}(f+f^{\dagger})d_{\sigma}+\eta^{*}_{\sigma}d_{\sigma}^{\dagger}(f+f^{\dagger}),
\end{eqnarray}
where $t_{\textbf{k}}$ are the hopping integrals between the QD and the normal lead, and $\eta_{\sigma}$ are the hopping
integrals
between the QD and the  Majorana lead. In terms of the phase $u_\s$, they are simply given as
$\eta_{\sigma}=\sqrt{2}\,\eta\, u_{\sigma}$~\cite{Leijnse.11}. In our configuration we consider the 
strongly anisotropic situation, corresponding to $\eta\ll t_{\mathbf k}$.
As we are interested in the limiting case in which the dot level is singly occupied, $\langle n_d\rangle \simeq 1$, 
the Anderson model at small energies is mapped onto the Kondo model by the Schrieffer-Wolff transformations. 
The so called local moment regime is formed by projecting out the empty and the doubly occupied states 
in the original SIAM. Folowing Hewson~\cite{Hewson}, we introduce the projector  
$P_n$ onto the subspace with $n=\{0, 1, 2\}$ particles, and write the total Hamiltonian of the system
as $H=\sum_{n,m}P_nHP_m$. 
Denoting $H_{nm}=P_nHP_m$, the effective Hamiltonian for $n=1$ is 
\begin{equation}\label{eq:sw}
H_{SW}=H_{11}+\sum_{n=0,2} H_{1n}(E-H_{nn})^{-1}H_{n1}\, 
\end{equation}
with $E$ the eigen-energy of $H_{SW}$. The sum in~\eqref{eq:sw} consists of two terms
corresponding to the empty ($n=0$) and doubly occupied ($n=2$) states.
Explicit evaluation of one of the terms, for example the one 
corresponding to $n=2$, gives:
\begin{eqnarray}\label{eq:h12}
&&H_{12}\frac{1}{E-H_{22}}H_{12}=\nonumber\\
&&=\sum_{\substack{\textbf{k} \textbf{k'}\\ \sigma,\sigma'}}\frac{t_{\textbf {k}}t^*_{\textbf{k'}}}{E-(2\epsilon_d+U)-H_{00}}\;c_{\textbf{k}\sigma}^{\dagger}n_{\bar{\sigma}}d_{\sigma}d^{\dagger}_{\sigma'}n_{\bar{\sigma'}}
c_{\textbf{k}'\sigma'}+\nonumber\\
&&+\sum_{\substack{\textbf{k}, \sigma,\sigma'}}\frac{t_{\textbf {k}}\eta^*_{\sigma'}}{E-(2\epsilon_d+U)
-H_{00}}\;c_{\textbf{k}\sigma}^{\dagger}n_{\bar{\sigma}}d_{\sigma}d^{\dagger}_{\sigma'}n_{\bar{\sigma'}}
(f+f^{\dagger})+\nonumber\\
&&+\sum_{\substack{\textbf{k'}, \sigma,\sigma'}}\frac{\eta_{\sigma}t^*_{\textbf{k'}}}{E-(2\epsilon_d+U)
-H_{00}}\;(f+f^{\dagger})n_{\bar{\sigma}}d_{\sigma}d^{\dagger}_{\sigma'}n_{\bar{\sigma'}}
c_{\textbf{k}'\sigma'},\nonumber\\
\end{eqnarray}
with the observation that $\{\xi,\epsilon_{\textbf{k}}\} \ll U$ is always satisfied in local moment regime. 
The first term in Eq. \eqref{eq:h12} can be further transformed along the lines of Ref.~\cite{Schrieffer.66}
into the "regular" Kondo Hamiltonian, while the second and third terms  entail the 
Majorana modes. The anisotropic situation that we are interested in allows us to 
neglect in  \eqref{eq:h12} terms $\sim \eta^2$. 
Spin-flip and potential scattering can be separated as follows by a simple trick that uses a identity for the Pauli matrices  
(we transform here the third term in Eq.~\eqref{eq:h12} only)
\begin{eqnarray}
&&\sum_{\substack{\textbf{k'}, \sigma,\sigma'}}\frac{\eta_{\sigma}t^*_{\textbf{k}'}}{E-(2\epsilon_d+U)
-H_{00}}\;(f+f^{\dagger})n_{\bar{\sigma}}d_{\sigma}d^{\dagger}_{\sigma'}n_{\bar{\sigma'}}
c_{\textbf{k}'\sigma'}\simeq\nonumber\\
&&\simeq -\sum_{\substack{\textbf{k'}, \sigma,\sigma'}}\frac{\sqrt{2}\eta t^*_{\textbf{k}'}}{U+\epsilon_d}\;
u_{\sigma}(f+f^{\dagger})d_{\sigma}d^{\dagger}_{\sigma'}=\nonumber\\
&&= 2\sum_{\substack{\textbf{k'}, \sigma,\sigma'\\\alpha,\alpha'}}\frac{\sqrt{2}\eta t^*_{\textbf{k}'}}{U+\epsilon_d}\times\;\nonumber\\
&&\left\{\left [ u_{\sigma}(f+f^{\dagger})\left(\frac{1}{2}\boldsymbol{\s}_{\sigma \sigma'}\right)c_{\textbf{k}'\sigma'}\right ]
\left [d_{\alpha'}^{\dagger}\left(\frac{1}{2}\boldsymbol{\s}_{\alpha \alpha'}\right)d_{\alpha} \right ]\right.+\nonumber\\
&&+\left.\left [ u_{\sigma}(f+f^{\dagger})\left(\frac{1}{2}\delta_{\sigma \sigma'}\right)c_{\textbf{k}'\sigma'}\right ]
\left [d_{\alpha'}^{\dagger}\left(\frac{1}{2}\delta_{\alpha \alpha'}\right)d_{\alpha} \right ]\right\},\nonumber
\end{eqnarray}
Keeping in mind that  we are interested only in spin flipping processes, and neglecting in what follows the potential 
scattering terms, we recover the Hamiltonian~\eqref{eq:H_int}. Neglecting also the momentum dependence, 
the exchange interactions in Eq.~\eqref{eq:H_int} are simply given by
\begin{eqnarray}
&&J=2 \;|t|^2\left \{ \frac{1}{U+\epsilon_d}-
\frac{1}{\epsilon_d}\right \},\\
&&J_{M}=2\,\sqrt{2}\;\eta t^*\left \{
\frac{1}{U+\epsilon_d}-
\frac{1}{\epsilon_d}\right \}.
\end{eqnarray}

As $\eta\ll t$, it immediately implies that $J_{M}\ll J$. Although it may be small, the exchange interaction $J_M$
couples the local spin to the Majorana modes and is responsible for transport across the dot.

\section{Green's functions} \label{app:SF}
In general, the time ordered, anti-time ordered, bigger and lesser Green's functions for an operator ${\cal O}_{\s}(t)$ in the Heisenberg representation, 
are defined respectively as:
\begin{eqnarray}
&&g_{{\cal O};\sigma}^t(t)=-i\left \langle {\cal T} {\cal O}_{\s}(t){\cal O}_{\s}^{\dagger}(0) \right \rangle,\\ 
&&g_{{\cal O};\sigma}^{\tilde t}(t)=-i\left \langle  {\tilde{\cal T} \cal O}_{\s}(t),{\cal O}_{\s}^{\dagger}(0) 
\right \rangle,\\
&&g_{{\cal O};\sigma}^>(t)=-i\left \langle  {\cal O}_{\s}(t){\cal O}_{\s}^{\dagger}(0) 
\right \rangle,\\
&&g_{{\cal O};\sigma}^<(t)=i\left \langle {\cal O}_{\s}^{\dagger}(0) {\cal O}_{\s}(t) 
\right \rangle.
\end{eqnarray}
The frequency dependent Green's functions are obtained by using the Fourier transform:
\begin{eqnarray}
g_{{\cal O};\sigma}^{t,\tilde t,>,<}(\omega)=\int_{-\infty}^{\infty} e^{i \omega t}g_{{\cal O};\sigma}^{t,\tilde t,>,<}(t)\; dt.
\end{eqnarray}
Then, the retarded Green's function $g_{{\cal O};\sigma}^R(\omega)=\Theta(t)\left [ g_{{\cal O};\sigma}^>(t)-g_{{\cal O};\sigma}^<(t)\right ]$, 
and the corresponding spectral function is defined as $A_{\cal O}^{\sigma}(\omega)=-1/\pi\; \textrm{Im} g_{{\cal O};\sigma}^R(\omega)$.

Next, we provide  a list with the non-interaction Green's functions used in the calculations. 
The Majorana Green's functions in the time domain are given by
\begin{eqnarray}
i\; \Xi_{\gamma_1}^>(t)&=& \frac{1}{2}\left[f(\xi)e^{i(\xi+\mu_R)t}+f(-\xi)e^{-i(\xi+\mu_R)t} \right],\nonumber\\
-i\;\Xi_{\gamma_1}^<(t)&=& \frac{1}{2}\left[f(\xi)e^{-i(\xi+\mu_R)t}+f(-\xi)e^{i(\xi+\mu_R)t} \right],\nonumber\\
i\;\Xi_{\gamma_1}^t(t)&=& \frac{\sgn(t)}{2}\left[f(\xi)e^{i(\xi+\mu_R)|t|}+f(-\xi)e^{-i(\xi+\mu_R)|t|} \right],\nonumber\\
-i\;\Xi_{\gamma_1}^{\tilde t}(t)&=& \frac{\sgn(t)}{2}\left[f(\xi)e^{-i(\xi+\mu_R)|t|}+f(-\xi)e^{i(\xi+\mu_R)|t|} \right].\nonumber\\
\end{eqnarray}
Here, we  denote by  $\mu_{L}/\mu_{R}$ the left/right lead chemical potentials
and $f(\omega)$ is the Fermi function.
The conduction electrons' Green's functions are
\begin{eqnarray}
i{\cal G}_{\sigma}^>(t)&=&\frac{\varrho_0}{it+a}e^{-i\mu_L t},\\
-i{\cal G}_{\sigma}^<(t)&=&\frac{\varrho_0}{-it-a}e^{-i\mu_L t},\\
i{\cal G}_{\sigma}^t(t)&=&\frac{\varrho_0}{it+a\;\sgn(t)}e^{-i\mu_L t},\\
-i{\cal G}_{\sigma}^{\tilde t}(t)&=&\frac{\varrho_0}{-it+a\;\sgn(t)}e^{-i\mu_L t},
\end{eqnarray}
where $a$ is the inverse bandwidth of the conduction electrons, which are assumed to have an 
exponentially decreasing density of states $e^{-a\;|\omega|}$.

\bibliography{references}

\end{document}